\documentstyle[jkas]{article}

\beginpage{1}
\endpage{4}
\year{2013}\volume{46}\month{February}

\runningauthor {H. Sung et al. (2013)} 
\runningtitle{Sejong Open cluster Survey (SOS)}

\month{June} \year{2013} \volume{46} \issueno{3}
\beginpage{97}\endpage{118}
\date{Received April 17, 2013; Accepted May 8, 2013}

\begin{document}
\title{SEJONG OPEN CLUSTER SURVEY (SOS). \\
0. TARGET SELECTION AND DATA ANALYSIS} 
\author{Hwankyung Sung$^{1}$, Beomdu Lim$^{1}$, Michael S. Bessell$^2$, Jinyoung S. Kim$^3$, Hyenoh Hur$^{1}$, Moo-Young Chun$^{4}$, and Byeong-Gon Park$^{4}$} 
\address{$^1$ Department of Astronomy and Space Science, Sejong University, 209,
Neungdong-ro, Gwangjin-gu, 
Seoul 143-747, Korea\\
 {\it E-mail : sungh@sejong.ac.kr}}
\address{$^2$ Research School of Astronomy and Astrophysics, Australian National University,
MSO, Cotter Road, Weston, ACT 2611, Australia \\ 
{\it E-mail : bessell@mso.anu.edu.au}}
\address{$^3$ Steward Observatory, University of Arizona,
933 N. Cherry Ave., Tucson, AZ 85721-0065, USA \\ 
{\it E-mail : serena@as.arizona.edu}}
\address{$^4$ Korea Astronomy and Space Science Institute,
776, Daedeok-daero, Yuseong-gu, Daejeon 305-348, Korea \\
{\it E-mail : mychun,bgpark@kasi.re.kr}}

\address{\normalsize{\it (Received April 17, 2013; Accepted May 8, 2013)}}
\offprints{H. Sung}
\abstract{
Star clusters are superb astrophysical laboratories containing
cospatial and coeval samples of stars with similar chemical composition.
We have initiated the Sejong Open cluster Survey (SOS) - a project dedicated 
to providing homogeneous photometry of a large number of open clusters in 
the SAAO Johnson-Cousins' $UBVI$ system. To achieve our main goal, we have paid much attention to 
the observation of standard stars in order to reproduce the SAAO standard system.\\
Many of our targets are relatively small, sparse clusters that escaped previous
observations. As clusters are considered building blocks of the Galactic disk,
their physical properties such as the initial mass function, the pattern
of mass segregation, etc. give valuable information on the formation and
evolution of the Galactic disk. The spatial distribution of young open
clusters will be used to revise the local spiral arm structure of the Galaxy.
In addition, the homogeneous data can also be used to test
stellar evolutionary theory, especially concerning rare massive stars.
In this paper we present the target selection criteria, the observational strategy
for accurate photometry, and the adopted calibrations for data analysis such as
color-color relations, zero-age main sequence relations, Sp - M$_V$ relations,
Sp - T$_{\rm eff}$ relations, Sp - color relations, and T$_{\rm eff}$ - BC relations.
Finally we provide some data analysis such as the determination of the
reddening law, the membership selection criteria, and distance determination.
}

\maketitle

\section{INTRODUCTION}

Open clusters are stellar systems containing a few (100 --  1000)
coeval stars with nearly the same chemical composition. They are
ideal targets to test stellar evolution theory.
As open clusters are stellar systems, they provide valuable information on
the distance and age of stars in the cluster, information that is very difficult
to obtain from field stars. In contrast to globular clusters,
open clusters have a wide range of ages. As they are important
building blocks of the Galactic disk, the distribution of age and abundance
of open clusters provide information on the star formation history
in the Galaxy. In addition, the stellar initial mass function (IMF) from open
clusters is one of the basic ingredients in constructing the star formation history
of the Galaxy as well as the population synthesis of unresolved remote galaxies.

The number of known open clusters is about 1700, but some seem to be
not real clusters \citep{cheon10} and some may be remnants of disrupted 
clusters. Owing to several all sky surveys such as {\it Hipparcos}, {\it Tycho},
2MASS, etc, many new open clusters or associations were identified 
\citep{bica03,dutra03,karachenko05}. Open clusters can be classified into 
three groups according to their age \citep{sung95}
-- young (age $\lesssim 10^7$ yrs), intermediate-age (age: $10^7$ -- $7 \times
10^8$ yrs), and old (age $\gtrsim 7 \times 10^8$ yrs).
Young open clusters can give information on the stellar evolution of
massive stars \citep{conti83,massey03,kook10,lim13} as well as on low-mass pre-main sequence (PMS) 
stars \citep{sung97,luhman12}. As massive
stars are still in the main sequence (MS) or in evolved stages, young open
clusters are ideal targets for studying the stellar IMF in a wide mass range 
\citep{sung04a,sung04b,sung10}. In addition, as they are still in or near
their birthplace, the spiral arm structure of the Galaxy can be derived
from the spatial distribution of young open clusters.
Typical young open clusters are the Trapezium cluster (or the Orion Nebula 
Cluster, ONC), NGC 2264 and NGC 2244 in Moncerotis, IC 1805 and 
IC 1848 in Cassiopeia, NGC 6231 in Scorpius, NGC 6530 (M8) and M20 (the Trifid 
nebula) in Sagittarius, NGC 6611 (the Eagle nebula) in Serpens, Trumpler 14 
(Tr 14), Tr 15, Tr 16, Collinder 228 (Cr 228), and Cr 232 in the $\eta$ Carina 
nebula. The double cluster {\it h} \& $\chi$ Per is slightly older than
the clusters listed above, but they are still classified as young open clusters.

As old open clusters are considered representative of the first generation of stars in
the Galactic disk, they can give information on the star formation history
in the Galactic disk and the chemical evolution of the Galaxy \citep{kim03}. As an open
cluster is a stellar system, stars in the cluster are subject to dynamical evolution \citep{sung99b}. 
As their age is older than the typical relaxation time scale of
open clusters, intermediate-age and old open clusters are good laboratories for
testing the dynamical evolution of multi-mass systems. Typical 
intermediate-age open clusters are the Pleiades and Hyades
in Taurus, Praesepe in Cancer, M35 in Gemini, and M11 in Scutum. 
Typical old clusters are M67, NGC 188, and NGC 6971.

\citet{lada03} and \citet{porras03} found that about 80\% of the stars in star forming
regions in the Solar neighborhood are in clusters with at least 100 members.
As small clusters or groups are dynamically unbound,
the stars in small clusters and groups will disperse and become
the field stars in the Galactic disk. Unfortunately, as most observations
are mainly focused on relatively rich open clusters, we have not much
information on these small clusters. In addition, the number of stars in
such small clusters is insufficient to test reliably stellar evolution
theory, as well as dynamical evolution models.
These obstacles can be overcome by combining data for several open clusters with
a similar age. This is the importance of the open cluster survey project.

Many photometric surveys of open clusters have been conducted up to now. Among them
``Photometry of stars in Galactic cluster field'' \citep{hoag61} was the first
comprehensive photoelectric and photographic photometry of open clusters in 
the Northern hemisphere. In the 1970s, N. Vogt and  A. F. J. Moffat performed $UBV$ \& H$\beta$ 
photoelectric photometry of many young open clusters in the Southern hemisphere 
\citep{vogt72,vogt73,moffat73a,moffat75a,moffat75b,moffat75c} and in the {\it l} = 
135$^\circ$ region \citep{moffat73b}. 

As the photon collecting area of CCDs increased rapidly in the 2000s, several 
photometric surveys of open clusters based on CCD photometry were started. 
Photometric surveys of open clusters in the 1990s were performed by several 
researchers (see \citet{ann99} for a summary). K. A. Janes and R. L. Phelps
performed photometric surveys of open clusters in the Northern hemisphere
\citep{janes93,phelps94}. \citet{ann99,ann02} 
started the BOAO Photometric Atlas of Open Clusters with the Bohyun-san 
Optical Astronomy Observatory (BOAO) 1.8m telescope to understand the structure 
of clusters and of the Galactic disk. They selected 343 target 
clusters and ambitiously started the project, but because of poor weather 
conditions in Korea, only two papers (photometric data for 16 clusters) were
published from the survey project. 
The CFHT Open Star Cluster Survey \citep{kalirai01a,kalirai01b,kalirai01c}
selected and observed 19 intermediate-age open clusters, but the survey team 
published deep photometric data for only 4 open clusters. The WIYN Open Cluster Study 
(WOCS) \citep{mathieu00} is the most successful open cluster survey program
up to now. The WOCS team selected cluster members from deep photometry, 
spectroscopic radial velocity surveys, and proper motion studies. They studied
binarity, stellar activity, chemical composition as well as observational
tests of stellar evolution theory. The Bologna Open Cluster Chemical Evolution 
Project \citep{bragaglia06} published photometric data for 16 open clusters
obtained with 1m to 4m-class telescopes. Recently \citet{maciejewski07}
published $BV$ photometric data for 42 open clusters, but their data are
very shallow due to the small aperture size of the telescope and shows
a large scatter probably due to bad weather.

We have started the Sejong Open cluster Survey (hereafter SOS) \citep{lim11}, a project 
dedicated to provide homogeneous photometry of a large number of open clusters
in the Johnson-Cousins' $UBVI$ system which is tightly matched to the SAAO 
standard $UBVRI$ system. To achieve our main goal we will pay much attention to 
the observation of standard stars in order to reproduce the SAAO standard system 
\citep{menzies89,menzies91,kilkenny98}. We have already derived the standard
transformation relations for the AZT-22 1.5m telescope at Maidanak Astronomical
Observatory (MAO) in Uzbekistan \citep{lim09}, for the Kuiper 61$''$ on Mount
Bigelow, Arizona, USA (Lim et al. 2013, in preparation), as well as for 
the 1m telescope at Siding Spring Observatory (SSO) \citep{sung00b}.

The homogeneous photometric data from this project can be used in
the study of \\
(1) the local spiral arm structure of the Galaxy \\
(2) the observational test of stellar evolution theory \\
(3) the stellar IMF \\
(4) the dynamical evolution of star clusters \\
(5) the star formation history of the Galaxy \\
(6) the chemical evolution of the Galaxy \\

In Section 2, we will describe the target selection criteria, and the spatial 
distribution of target clusters. 
In Section 3, we describe our strategy for accurate photometry
such as determination of the atmospheric extinction coefficients, transformation
coefficients, and correction for the other factors affecting photometry.
In Section 4, we adopt several calibrations required for the analysis of 
photometric data such as the intrinsic color relations, zero-age main sequence
(ZAMS) relations, the Spectral Type (Sp) - M$_V$ relation, the Sp - T$_{\rm eff}$ relation,
etc. In Section 5, we present some data analysis such as the determination
of the reddening law, the membership selection criteria, and distance 
determination. We summarize our main results in Section 6.

\section{TARGET SELECTION}

\subsection{Status of Open Cluster Data}

\begin{figure}[!b]
\centering \epsfxsize=8cm 
\epsfbox{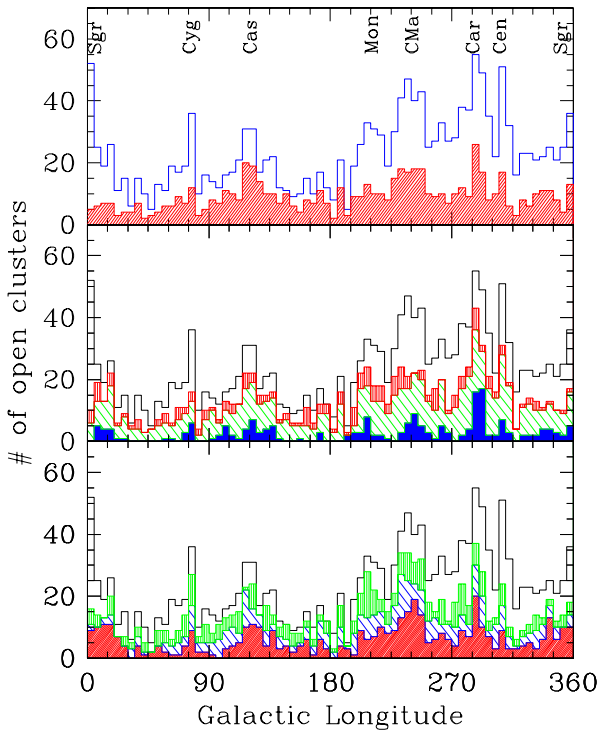} 
\caption{Distribution of open clusters with Galactic longitude.
         (Top) Open histogram: number of all open clusters at a given
         longitude bin. Red hatched histogram: number of open clusters
         observed in the $UBV$ photometric system.
         (Middle) Distribution of open clusters according to their age
         -- blue: young open clusters, green: intermediate-age open clusters,
         red: old open clusters, white: unknown.
         (Bottom) Distribution of open clusters according to the survey
         priority (see Section 2.2). -- red: priority 1, blue: priority 2, 
         green: priority 3, white: priority 4.}
\label{fig_glong}
\end{figure}

After the introduction of CCDs in astronomy, most observations were focused
on densely populated clusters, globular clusters, or external galaxies, and
the observational status of relatively poor open clusters was largely neglected. 
Among 1686 open clusters listed in the open cluster data 
base WEBDA\footnote{http://www.univie.ac.at/webda/}, about 550 clusters 
were observed either by means of $UBV$ photoelectric photometry or modern CCD 
photometry, i.e. two thirds of the listed open clusters have still not been observed 
even in the $UBV$ system. Firstly, we checked photometric data of all the open 
clusters in WEBDA. The spatial distribution and the observational status 
of open clusters versus Galactic longitude are shown in Figure \ref{fig_glong}.

\begin{figure}[!t]
\centering \epsfxsize=6.5cm 
\epsfbox{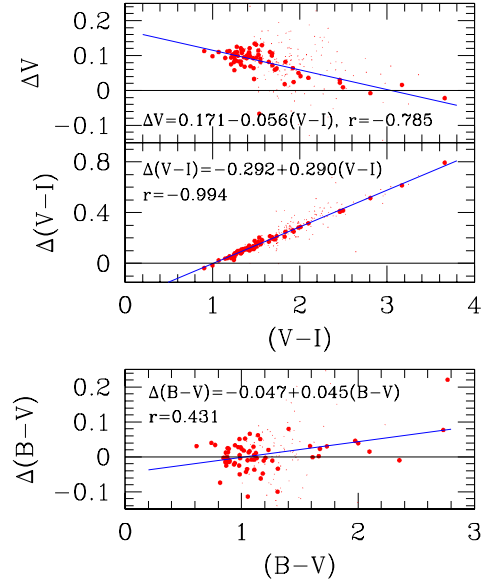} 
\caption{Comparison of the photometry presented by Kook et al. (2010)
         with that of \citet{piatti01} for the stars in the young
         open cluster Hogg 15. Large and small dots represent bright ($V \leq
         17$ mag) and faint stars ($V > 17$ mag), respectively.
         There are large systematic differences
         in $V$ and $V-I$ between Kook et al. (2010) and
         \citet{piatti01}, but the difference in $B-V$ is
         not so pronunced. \label{fig_delta} }
\end{figure}

Until now several open cluster survey projects have been performed or are
in progress. In order to derive meaningful results from the combined photometric 
data of open clusters with a similar age, the most important factor is the
homogeneity of the photometry. Unfortunately, some 
photometric data show large deviations (see \citet{mermilliod03} or 
Figure \ref{fig_delta}). The size of the deviations is much larger than the 
expected errors from uncertainty in atmospheric extinction correction. 
The large systematic differences shown in Figure \ref{fig_delta}
seem to be caused by errors, either in the atmospheric extinction
correction or in the transformation to the standard system, or both.

\begin{figure}[!t]
\centering
\epsfxsize=6cm \epsfbox{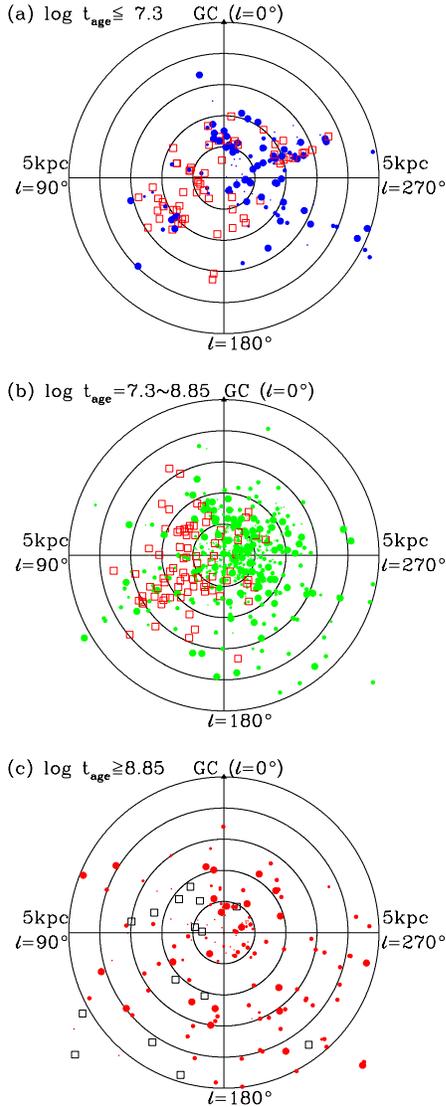} 
\caption{Distribution of target clusters projected on the Galactic 
          plane. The size of dots represents the priority of observation,
         and the square represents the open clusters already observed.
         (Upper) Distribution of young open clusters.
         (Middle) Distribution of intermediate-age open clusters.
         (Lower) Distribution of old open clusters. \label{fig_gdisk}}
\end{figure}

We started a project to observe a large number of open clusters
in both the Southern and northern hemispheres. We observed several
open clusters in the nouthern hemisphere with the 1m telescope at SSO,
and observed more open clusters with 
the 1m telescope at CTIO in 2011. In addition,
we have obtained the images of many Northern open clusters using the AZT-22 
1.5m telescope at MAO in Uzbekistan over 5 years from August 2004.
We discovered that the atmospheric extinction coefficients at MAO are
very high during the summer season due to dust from the desert. Those in fall
and winter are normal. We are observing more open clusters in the
northern hemisphere with the Kuiper 61$''$ telescope in Arizona, USA 
from October 2011.

\subsection{Target Selection Criteria}

We downloaded a $\sim 15' \times 15'$ image for most of the open clusters from
the STSci Digitized Sky Survey\footnote{http://archive.stsci.edu/cgi-bin/dss\_form}, and checked all the images visually.
The target selection criteria are \\
- priority 1 : young open clusters or dense open clusters with no $UBVI$ data \\
- priority 2 : relatively dense open clusters without $UBVI$ data or 
well-observed open clusters for the homogeneity of photometry \\
- priority 3 : sparse open clusters \\
- priority 4 : no clustering of stars.

We have selected 455, 318, 318, and 604 clusters with priority 1, 2, 3, and 4, 
respectively. Among them, 125 priority 1 targets, 47 priority 2 targets, 15 
priority 3 targets, and 10 priority 4 targets have been observed with the telescopes
listed above (up to February 2013). The spatial distributions of selected clusters
with Galactic longitude is in the bottom panel of Figure \ref{fig_glong}.
We also plotted the spatial distribution of open clusters projected onto the 
Galactic plane for three age groups in Figure \ref{fig_gdisk}. In the upper
panel, we limited the open clusters to those younger than 20 Myr because
the age of many young open clusters was assigned to be about 10 Myr.
The local spiral arm structure can clearly be seen in the upper
panel, and is barely visible in the middle panel. The old open clusters
are distributed evenly in the plane.
Currently more than half of the priority 1 targets in 
the Northern hemisphere have been observed, and we are now searching for 
a 1m-class telescope in the Southern hemisphere.

\section{OBSERVATION STRATEGY}

To perform accurate photometry, we should pay as much attention to the 
preparation of the observations as to the observation of standard stars.
It is well-known that there are some systematic differences between the
Landolt standard $UBVRI$ system \citep{landolt92} and the SAAO $UBVRI$ 
system \citep{menzies89,menzies91,bessell95}.
Hence, it is important to have some knowledge of the photometric system.
Without knowledge of the characteristics of an observing site and of the
photometric system, it is impossible to achieve 1\% error levels in photometry.

\subsection{Standard Stars}

The selection of standard stars is one of the most important aspects within
the standard system photometry.  We found the necessity of a non-linear 
correction term in the transformation
to the \citet{landolt92} $U$ system \citep{sung98,sung00b}. To avoid this and 
to obtain accurate standardized data in $U-B$, we observed the SAAO 
secondary standard Equatorial stars \citep{menzies91} and the extremely blue 
and red stars from \citet{kilkenny98}. Unfortunately, most of these stars are very bright,
and we have to use short exposure times for them.
We carefully check for uneven illumination patterns or systematic differences
in the effective exposure time for the images obtained with short exposures
($\tau_{exp} \lesssim 10$ s, see for example \citet{lim08}).

\subsection{Extinction Coefficients}

Atmospheric extinction is caused by absorption and scattering by air
molecules or other particles in the Earth's atmosphere.
Most of the extinction in the visual window is due to Rayleigh scattering 
($\propto { 1 \over \lambda^4}$)
by air molecules. Another important and variable contributor to the extinction is
the scattering and absorption by small liquid or solid particles of various
sizes called aerosols \citep{cousins98}.
The total extinction value depends primarily on the line-of-sight length through
the Earth's atmosphere (air mass). In addition, since
the extinction varies with wavelength, the mean value of the extinction measured
across a wide filter pass band will differ depending on the spectral energy
distribution of the stars. We correct for these effects by looking for
a primary or first extinction coefficient that depends on air mass
but is independent of color, and on a secondary extinction coefficient
that depends also on color. 

The magnitude corrected for atmospheric extinction is given by

\begin{equation}
m_{\lambda,0} = m_{\lambda}-(k_{1\lambda}-k_{2\lambda}   C)   X,
\end{equation}

\noindent
where $m_{\lambda,0}$, $m_{\lambda}$, $k_{1\lambda}$, $k_{2\lambda}$, $C$,
and $X$ are the extinction-corrected magnitude, observed instrumental
magnitude, primary extinction coefficient, secondary extinction coefficient,
relevant color index, and air mass, respectively. In general, we observe many
standard regions several times at various air masses. To get a long
baseline in air mass we often observe standard stars near the meridian
and again at zenith distances of $\approx60^\circ$. The secondary
extinction coefficients for $VRI$ are very small and normally ignored.

The atmospheric extinction coefficients at SSO are presented in \citet{sung00b},
and those at MAO in \citet{lim09}.
At MAO there are obvious seasonal variations in the primary extinction
coefficients for all filters. The coefficients are larger in summer, but
smaller in winter. In addition, the extinction coefficients in summer show
a large scatter. The mean extinction coefficients are slightly larger at MAO
in $V$ and $R$ than those at SSO, but those in $U$, $B$, and $I$ are slightly smaller
at MAO. In addition, we find about 10\% or more real fluctuations in the
extinction coefficients. It is better to determine the extinction coefficients
every night unless the standard stars and program objects are observed and
interspersed at similar zenith distances.

\subsection{Transformation Coefficients}

All photometric systems are defined by the filters and detectors used in
the observations.  Slight deviations between standard magnitudes
and atmospheric-extinction-corrected natural instrumental magnitudes are to be
expected, and need to be corrected for to achieve the highest accuracy.
Such differences are tracked and corrected through the observation of
many standard stars with the largest possible range of colors. The correction
terms between two systems are called transformation coefficients, and
are related as

\begin{equation}
M_{\lambda} = m_{\lambda,0}+\eta_{\lambda}  C+\zeta_{\lambda},
\end{equation}

\noindent
where $M_{\lambda}$, $m_{\lambda,0}$, $\eta_{\lambda}$, $C$,
$\zeta_{\lambda}$ represent the standard magnitude, atmospheric
extinction-corrected instrumental magnitude as defined in Equation (1),
transformation coefficient, relevant color index,
and photometric zero point, respectively.
Normally, the transformation relation against a relevant color is a single
straight line or a combination of several straight lines (see \citet{sung08a}).
The final transformation relations can be determined using all the standard
stars observed for several years after correcting for daily differences, such
as extinction and photometric zero points.

\subsection{Time Variation of Photometric Zero Points}

The photometric zero points depend primarily on the light gathering power
of the photometric system, i.e. the size and state of the primary mirror
and the quantum efficiency of the detector. In addition, changes in
atmospheric conditions such as a change in aerosol, water vapor or dust content
in the atmosphere or a variation in the ozone layer in the upper atmosphere
also affect the zero points. It is known that changes in water vapor
content affect the extinction at longer wavelengths (mostly $I$ and
near-IR), while changes in the aerosol content  affect the extinction at all optical wavelengths.
Changes in the ozone layer mainly affect $U$, $V$ and $R$. At MAO the time
variation in many cases started at evening twilight and ended around midnight
\citep{lim09}. Such a variation at MAO may be related to the change
in the content of water vapor.

\subsection{Spatial Variation}

In some cases we have to consider a spatial term in the transformation
relations due to the uneven illumination of the focal plane. We initially 
tried to determine standard transformation relations for the CFH12K CCD of
CFHT using Stetson's extensive photometry of standard star regions
\citep{stetson00}, but we were unable to do so with a reasonable error.
Later, we found that a star's position on the chip influenced the
transformation \citep{sung08a}. Such an effect has now been identified on all
wide-field imagers and results from the non-uniform illumination of
the mosaic plane that cannot be corrected by normal flat field exposures,
due to scattered light. Recently we also found similar corrections for 
the CTIO 4m MOSAIC II CCDs \citep{lim13}.

\subsection{General Form of the Transformation Relation}

Now we can write a general form of the extinction and transformation relation
as follow (see \citet{sung08a}).

\begin{equation}
M_\lambda = m_{\lambda, 0}
+ \eta_\lambda   C + \alpha_\lambda   \hat{UT} 
+ \beta_\lambda   \hat{x}_{CCD}
+ \gamma_\lambda   \hat{y}_{CCD} + \zeta_\lambda
\end{equation}

\noindent
where $\alpha_\lambda$, $\hat{UT}$, $\beta_\lambda$, $\gamma_\lambda$, 
$\hat{x}_{CCD}$, and $\hat{y}_{CCD}$ denote the time-variation coefficient, 
the time difference relative to midnight, the spatial variation coefficient in 
the x- and y- coordinate, CCD x- and y-coordinates in units of 1,000 pixels, 
respectively.
If the CCD used in the observations is a 2K-single chip, the spatial term may 
also be neglected. We found a non-negligible spatial term in the
transformation of the Kuiper 61$''$ telescope.

\subsection{Sequence of Observations}

The extinction coefficients differ from night to night and often within a night.
Therefore, it would be better to observe as many standard regions/stars as possible to determine
the extinction coefficients as well as the time variation of the photometric
zero points (see \citet{sung00b} or \citet{lim09}).

We observe three or four standard stars/regions at various air masses just 
after the evening twilight and midnight, and observe one or more standard 
stars/regions just before the morning twilight. The observations near the 
evening twilight or midnight give the instantaneous extinction coefficients.
A temporary value of the transformation coefficients for the $VRI$ filters
($\eta_\lambda   C + \zeta_\lambda '$) can be determined from the plot
of $M_\lambda-m_\lambda$ against the relevant color $C$. For $U$ or $B$
the slope obtained above is $k_{2\lambda}   X + \eta_\lambda$. By applying
these temporary values for extinction and transformation coefficients
to all the data observed at different times we can calculate the time variation
of the photometric zero points. After correcting for the time variation, we can
calculate the transformation coefficients using all the data observed during
the night. Then, we recalculate the extinction coefficient using all
the data.

\section {ADOPTED CALIBRATIONS}

To analyze the photometric data from this survey, we have to use various
relations. Many investigators studied and adopted various relations used in
the data analysis such as the ZAMS relation, the color - temperature 
relation, the Sp - temperature relation, etc. In many cases, we are confronted 
with the situation in which the physical quantity determined from one 
relation differs from that derived from the other relations, e.g. the effective
temperature (T$_{\rm eff}$) of a star from the T$_{\rm eff}$ versus $B-V$ 
relation, and that from the T$_{\rm eff}$ versus $U-B$ relation. Such a 
discrepancy may be caused by the different resolution at a given T$_{\rm eff}$ 
range between colors, but in some cases it is caused by {\bf the lack
of internal consistency between relations}.

In this section we describe all the relations used in the data analysis. To get
reliable and self-consistent quantities, we choose the {\bf spectral type (Sp)}
as the primary calibrator. In addition, to derive internally consistent 
relations, we make use of the intrinsic color relations between colors
because such relations are relatively well known. Therefore, we first
derive the intrinsic color relations in Section 4.1. The other basic relations,
such as the ZAMS relation and the Sp - $M_V$ relations, are presented in 
the subsequent sections. We then adopt the Sp - T$_{\rm eff}$ relation, 
the Sp - color relation, and the T$_{\rm eff}$ - bolometric correction (BC) 
relation for a given luminosity class (LC) from Section 4.5. 

\subsection{Intrinsic Color Relations}

\begin{table}[!t]
\begin{center}
\centering
\caption{The Intrinsic Color Relations for Optical Colors \label{tab_cc}}
\doublerulesep2.0pt
\renewcommand\arraystretch{1.07}
\begin{tabular}{c|cccc|cc}
\hline \hline
 $B-V$ & \multicolumn{4}{c}{$U-B$} & \multicolumn{2}{c}{$V-I$}  \\ \hline
 L. C. &   V   &  III &  Ib &  Ia  &    V    & III  \\ \hline
 -.325 & -1.18 & -1.20 & -1.18 & -1.18 & -.355 & -0.34 \\
 -0.30 & -1.06 & -1.10 & -1.15 & -1.15 & -0.33 & -0.32 \\
 -.275 & -0.99 & -1.00 & -1.12 & -1.12 & -.305 & -.304 \\
 -0.25 & -0.89 & -0.90 & -1.08 & -1.09 & -0.28 & -0.27 \\
 -.225 & -0.79 & -0.79 & -1.04 & -1.05 & -0.25 & -0.23 \\
 -0.20 & -0.70 & -0.68 & -1.00 & -1.01 & -0.22 & -0.20 \\
 -.175 & -0.60 & -0.57 & -0.95 & -0.97 & -0.19 & -0.17 \\
 -0.15 & -0.50 & -0.47 & -0.88 & -0.92 & -0.16 & -0.14 \\
 -.125 & -0.40 & -0.37 & -0.80 & -0.86 & -0.13 & -0.11 \\
 -0.10 & -0.30 & -0.28 & -0.74 & -0.81 & -0.11 & -0.08 \\
 -.075 & -0.20 & -0.20 & -0.66 & -0.76 & -0.08 & -.053 \\
 -0.05 & -0.11 & -0.11 & -0.58 & -0.71 & -.056 & -.026 \\
 -.025 & -0.05 & -.055 & -0.48 & -0.65 & -0.03 &  0.00 \\
  0.00 &  0.00 &  0.00 & -0.35 & -0.59 & -.003 &  .027 \\
  0.05 &  .055 &  .065 & -0.15 & -0.33 &  .054 &  0.08 \\
  0.10 &  .085 &  0.11 &  0.01 & -0.06 &  0.11 &  0.13 \\
  0.15 &  .095 &  0.12 &  0.08 &  0.06 &  0.17 &  0.19 \\
  0.20 &  .085 &  0.11 &  0.16 &  0.15 &  0.23 &  .244 \\
  0.25 &  0.06 &  .095 &  0.23 &  0.23 &  .294 &  0.30 \\
  0.30 &  .035 &  .075 &  0.27 &  .275 &  .353 &  .356 \\
  0.35 &  .005 &  0.07 &  0.30 &  0.30 &  0.41 &  0.41 \\
  0.40 & -0.01 &  0.07 &  .325 &  0.33 &  0.47 &  .467 \\
  0.45 & -0.02 &  .075 &  0.35 &  0.35 &  .525 &  0.52 \\
  0.50 &  0.00 &  0.09 &  .375 &  .375 &  .575 &  0.57 \\
  0.60 &  0.09 &  0.16 &  0.42 &  0.42 &  0.66 &  0.67 \\
  0.70 &  0.23 &  0.28 &  0.45 &  0.45 &  0.74 &  0.76 \\
  0.80 &  0.41 &  0.44 &  0.49 &  0.49 &  0.84 &  0.83 \\
  0.90 &  0.65 &  0.62 &  0.63 &  0.63 &  0.95 &  0.89 \\
  1.00 &  0.86 &  0.84 &  0.81 &  0.81 &  1.08 &  0.96 \\
  1.10 &  1.04 &  1.04 &  0.98 &  0.98 &  1.22 &  1.045 \\
  1.20 &  1.13 &  1.24 &  1.15 &  1.15 &  1.375 &  1.14 \\
  1.30 &  1.20 &  1.44 &  1.33 &  1.33 &  1.555 &  1.253\\
  1.40 &  1.22 &  1.64 &  1.51 &  1.51 &  1.775 &  1.386\\
  1.50 &  1.17 &  1.81 &  1.68 &  1.68 &  2.25 &  1.57 \\
  1.60 &  1.19 &  1.89 &  1.86 &  1.86 &  2.60 &  1.80 \\ \hline
\end{tabular}
\end{center}
\end{table}

\begin{figure}[!t]
\centering
\epsfxsize=8cm \epsfbox{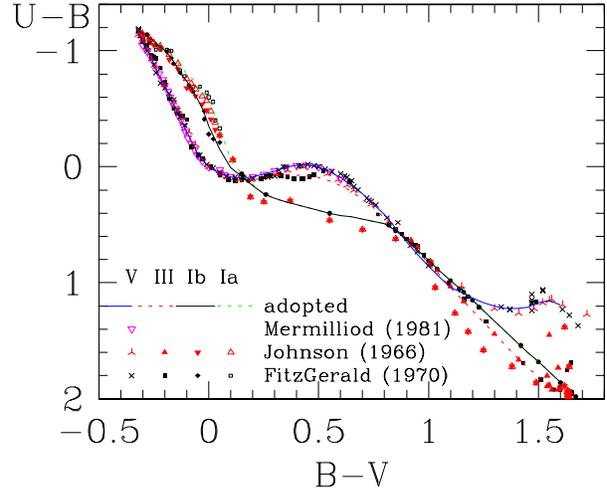} 
\caption{The ($U-B,~B-V$) color-color diagram. Solid and dashed lines 
         represent the adopted relations. The meaning of the symbols is 
         explained in the figure. \label{fig_cc}}
\end{figure}

The interstellar reddening can be determined from the two-color diagrams 
(hereinafter TCDs) using the difference in the color excess ratios between
colors. Although several TCDs are used in the reddening estimate,
the ($U-B,~B-V$) TCD is the most popular and a well-established one.
To derive the reddening $E(B-V)$, we should adopt two relations:
(1) the intrinsic color relation for the MS (or ZAMS) stars
in the ($U-B, ~ B-V$) diagram \footnote{The intrinsic color-color
relations for giants or supergiants may be used, but there are
several uncertainties such as the uncertainty in LC or in the intrinsic color. 
Therefore, the reliability of $E(B-V)$ from evolved stars is relatively poor.},
and (2) the slope of the reddening vector in the TCD.

The intrinsic color relation in the ($U-B, ~ B-V$) diagram was
investigated by several researchers after the introduction of the
$UBV$ photometric system. Among them, the relations derived by 
\citet{johnson66}, \citet{mermilliod81}, \citet{schmidt-kaler82}, and 
\citet{fitzgerald70} are most frequently used. We compared their 
relations in Figure \ref{fig_cc} and presented the adopted relation in Table
\ref{tab_cc}. While \citet{johnson66} and \citet{fitzgerald70} derived and
presented the relations for all the luminosity classes, \citet{mermilliod81}
presented the relation only for MS stars derived from a comparative
study of several well-observed open clusters. For early-type stars
we adopt mainly the relation presented by \citet{mermilliod81} for MS stars 
and the one of \citet{johnson66} for the other LCs. For late-type stars
we adopt the data by \citet{fitzgerald70} and smoothed the relations.

\begin{figure}[!t]
\centering
\epsfxsize=8cm \epsfbox{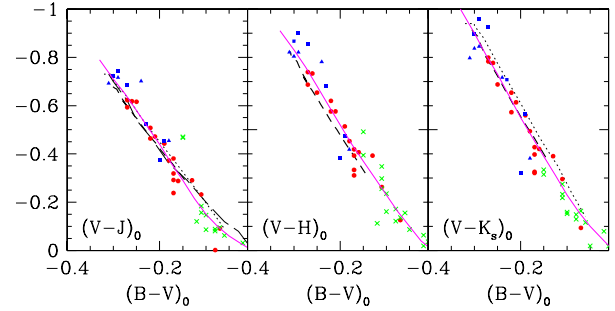} 
\caption{The intrinsic color relation between ($B-V$)$_0$ and 2MASS colors.
         Red dot, green cross, blue square and blue triangle represent
         respectively, the star in NGC 2362, the Pleiades, NGC 2264, and
         the ONC, while the thick dashed, thin long dashed, dotted, and solid
	 lines denote the synthetic color from Tlusty non-LTE model atmosphere,
    	 \citet{johnson66}'s relation for supergiants, \citet{johnson66}'s
	 relation for MS, and the adopted relation, respectively. \label{fig_nir}}
\end{figure}

%
The color excess in other colors, for e.g. $E(V-I)$, $E(V-J)$,
$E(V-H)$, or $E(V-K_s)$, cannot be easily determined from the TCD.
The color excess can be determined from
the intrinsic color relations between color and ($B-V$)$_0$.
In addition, as the near- and mid-infrared color excess ratios are very
sensitive to the reddening law \citep{guetter89}, i.e. the total
to selective extinction ratio $R_V \equiv A_V / E(B-V)$, we derived
the intrinsic color relations between ($B-V$)$_0$ and ($V-\lambda$)$_0$
for the 2MASS pass bands \citep{2mass}.

To derive those relations, we chose 2MASS data
for the Pleiades, NGC 2362, NGC 2264, and the ONC. 
In addition, we also calculated the synthetic colors in 2MASS and {\it Spitzer}
IRAC bands using the Tlusty non-LTE model atmosphere of O- and B-type stars
\citep{lanz03,lanz07}
\footnote{http://nova.astro.umd.edu/Tlusty2002/tlusty-frames-models.html}
by one of the authors (MSB). As these clusters are relatively
less reddened, the uncertainty due to the reddening correction can be
minimized. We assumed $E(B-V)$ = 0.10 and 0.04 mag for NGC 2362 and the 
Pleiades, respectively. The $E(B-V)$ of individual stars is estimated
and corrected for the stars in NGC 2264 and the ONC. In correcting for the color
excess for each color, we used the relation between the color excess ratio
and the value of $R_V$ of \citet{guetter89} for 2MASS colors.
We present the intrinsic color relations
in Figure \ref{fig_nir} 
and Table \ref{tab_ir}.

\begin{table}[!t]
\begin{center}
\centering
\caption{Intrinsic color relations for 2MASS colors
\label{tab_ir} }
\doublerulesep2.0pt
\renewcommand\arraystretch{1.07}
\begin{tabular}{c|ccc}
\hline \hline
$B-V$ & $V-J$ & $V-H$ & $V-Ks$ \\ \hline
-0.33 & -0.79 & -0.91 & -1.00 \\
-0.30 & -0.71 & -0.83 & -0.90 \\
-.275 & -0.65 & -0.76 & -0.81 \\
-0.25 & -0.58 & -0.68 & -0.73 \\
-.225 & -0.51 & -0.60 & -0.64 \\
-0.20 & -0.44 & -0.52 & -0.55 \\
-.175 & -0.36 & -0.44 & -0.47 \\
-0.15 & -0.29 & -0.37 & -0.39 \\
-.125 & -.215 & -0.30 & -0.31 \\
-0.10 & -0.15 & -0.23 & -0.23 \\
-.075 & -0.10 & -0.16 & -0.16 \\
-0.05 & -0.06 & -0.10 & -0.10 \\
-.025 & -0.03 & -0.04 & -0.05 \\
 0.00 &  0.00 &  0.00 &  0.00 \\ \hline
\end{tabular}
\end{center}
\end{table}

\begin{figure}[!b]
\centering
\epsfxsize=8cm \epsfbox{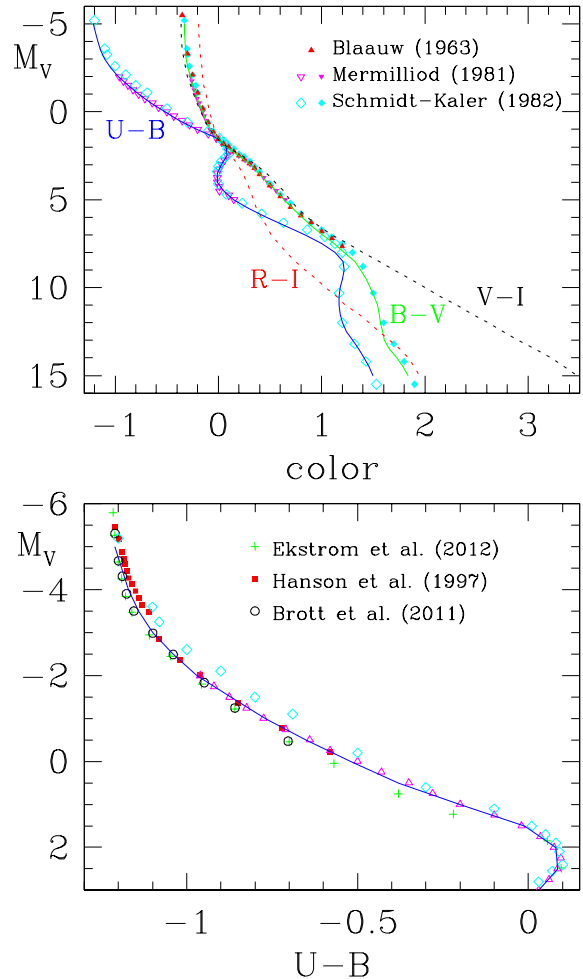} 
\caption{Zero-Age Main Sequence relations.
         The solid and dashed lines represent the adopted ZAMS relations for
         the given color indices. Open circles, open diamonds and triangles 
         represent the ZAMS relations by \citet{mermilliod81}, 
         \citet{schmidt-kaler82}, and \citet{blaauw63}, respectively.  
         \label{fig_zams}}
\end{figure}

\subsection{Zero-Age Main-Sequence Relations}

\begin{table}[!t]
\begin{center}
\centering
\caption{Zero-Age Main Sequence relations \label{tab_zams} }
\doublerulesep2.0pt
\renewcommand\arraystretch{1.07}
\begin{tabular}{c|cccc}
\hline \hline
  $M_V$ & $B-V$ & $U-B$ & $V-I$ & $R-I$ \\ \hline
  -5.0 & -0.33 & -1.21 & -0.36 & -0.19 \\
  -4.5 & -0.33 & -1.19 & -0.36 & -0.19 \\
  -4.0 & -.325 & -1.17 & -.355 & -0.18 \\
  -3.5 & -0.32 & -1.15 & -0.35 & -0.18 \\
  -3.0 & -0.31 & -1.10 & -0.33 & -0.17 \\
  -2.5 & -.295 & -1.04 & -0.32 & -0.16 \\
  -2.0 & -0.27 & -0.97 & -0.30 & -0.15 \\
  -1.5 & -.245 & -0.87 & -0.27 & -0.14 \\
  -1.0 & -0.22 & -0.77 & -0.24 & -0.12 \\
  -0.5 & -.185 & -0.65 & -.205 & -.105 \\
   0.0 & -.155 & -0.52 & -0.17 & -.085 \\
   0.5 & -0.12 & -0.38 & -0.13 & -0.06 \\
   1.0 & -.075 & -0.20 & -0.08 & -0.04 \\
   1.5 & -0.01 & -0.01 & -.015 & -.005 \\
   2.0 &  .085 &  0.08 &  0.09 &  0.06 \\
   2.5 &  0.20 &  .085 &  0.23 &  0.13 \\
   3.0 &  0.31 &  0.03 &  .365 &  0.19 \\
   3.5 &  .395 & -0.01 &  .465 &  .235 \\
   4.0 &  .475 & -0.01 &  0.55 &  .275 \\
   4.5 &  0.56 &  0.05 &  0.63 &  0.31 \\
   5.0 &  0.64 &  0.15 &  0.69 &  0.34 \\
   5.5 &  0.72 &  0.29 &  0.77 &  0.37 \\
   6.0 &  0.81 &  0.47 &  0.86 &  0.42 \\
   6.5 &  0.90 &  0.65 &  0.97 &  0.47 \\
   7.0 &  1.01 &  0.84 &  1.10 &  0.51 \\
   7.5 &  1.12 &  1.00 &  1.25 &  0.58 \\
   8.0 &  1.22 &  1.13 &  1.40 &  0.66 \\
   8.5 &  1.32 &  1.21 &  1.56 &  .745 \\
   9.0 &  1.38 &  1.22 &  1.70 &  0.84 \\
   9.5 &  1.43 &  1.20 &  1.86 &  0.95 \\
  10.0 &  1.47 &  1.18 &  2.00 &  1.04 \\
  11.0 &  1.54 &  1.17 &  2.31 &  1.23 \\
  12.0 &  1.57 &  1.20 &  2.61 &  1.46 \\
  13.0 &  1.61 &  1.30 &  2.90 &  1.66 \\
  14.0 &  1.73 &  1.41 &  3.22 &  1.84 \\
  15.0 &  1.84 &  1.50 &  3.48 &  1.95 \\ \hline
\end{tabular}
\end{center}
\end{table}

The ZAMS relation is the basic tool used to estimate
the distance to open clusters. As noticed by \citet{johnson56}, the distance to
an open cluster may have a large error if the effect of evolution during 
the MS phase is neglected. They introduced the standard main sequence
for age ``zero''. \citet{sandage57} used the term ``zero-age main sequence''. 
There are several standard clusters used
in deriving the ZAMS relation. The primary cluster is the Hyades, which was
the only cluster having a reliable distance at that time. The Pleiades is
the second cluster which can provide the ZAMS relation up to A-type stars.
Unfortunately there are no young open clusters within 1kpc from the Sun
(apart for some unsuitable extremely young clusters), which makes it difficult 
to extend the relation to the upper MS. The young open clusters are, in 
general, highly reddened, show a differential
reddening across the cluster, and have an anomalous reddening law in many cases.
In addition, the metallicity of stars in the Perseus arm is known to be
lower than that of the stars in the Solar neighborhood. The ZAMS relation of 
\citet{schmidt-kaler82} is nearly identical to that of \citet{blaauw63} which 
is very similar to that of \citet{johnson56} or \citet{sandage57}. 
\citet{mermilliod81} published a new ZAMS relation from the analysis of 
photometric data for many open clusters. His ZAMS relation is slightly 
fainter than the others \citep{blaauw63,schmidt-kaler82}.

\citet{sung99a} presented the ZAMS relation in $V-I$ color which is less 
sensitive to the metallicity difference. We adopt the ZAMS relations 
used in the data analysis. The upper part of the ZAMS is taken from the 
reddening corrected color-magnitude diagrams (CMDs) of the young open clusters 
in the $\eta$ Carina nebula \citep{hur12} and NGC 6611. By the definition of 
``zero'' age for single stars, we took {\bf the lower ridge line of the MS 
band as the ZAMS}. For B- to G-type stars, we adopt the ZAMS relation of 
\citet{mermilliod81}. For faint stars we adopted the relation of
\citet{schmidt-kaler82}. The ZAMS relations for $V-I$ and
$R-I$ are derived using the intrinsic color relations in Table \ref{tab_cc}.

\begin{table}[!h]
\begin{center}
\centering
\caption{Spectral type-$M_V$ relation \label{tab_sp_mv} }
\doublerulesep2.0pt
\renewcommand\arraystretch{1.07}
\begin{tabular}{c|cccccc}
\hline \hline
Sp& V & III & II & Ib & Iab & Ia \\ \hline
 O3 & -5.45 & -5.90 & -5.95 & -6.05 & -6.25 & -6.45 \\
 O4 & -5.35 & -5.85 & -5.95 & -6.05 & -6.35 & -6.60 \\
 O5 & -5.25 & -5.80 & -5.90 & -6.05 & -6.40 & -6.75 \\
 O6 & -5.10 & -5.70 & -5.85 & -6.05 & -6.45 & -6.85 \\
 O7 & -4.90 & -5.55 & -5.80 & -6.05 & -6.45 & -6.95 \\
 O8 & -4.70 & -5.35 & -5.70 & -6.00 & -6.50 & -7.00 \\
 O9 & -4.40 & -5.10 & -5.55 & -6.00 & -6.50 & -7.00 \\
 B0 & -3.85 & -4.70 & -5.35 & -5.95 & -6.50 & -7.05 \\
 B1 & -3.20 & -4.20 & -5.20 & -5.90 & -6.50 & -7.05 \\
 B2 & -2.50 & -3.60 & -5.00 & -5.85 & -6.50 & -7.05 \\
 B3 & -1.70 & -3.00 & -4.80 & -5.80 & -6.50 & -7.05 \\
 B4 & -1.25 & -2.55 & -4.60 & -5.75 & -6.50 & -7.05 \\
 B5 & -1.00 & -2.15 & -4.40 & -5.70 & -6.50 & -7.05 \\
 B6 & -0.70 & -1.85 & -4.20 & -5.65 & -6.50 & -7.05 \\
 B7 & -0.40 & -1.50 & -3.95 & -5.55 & -6.50 & -7.07 \\
 B8 & -0.15 & -1.20 & -3.65 & -5.50 & -6.50 & -7.10 \\
 B9 &  0.30 & -0.90 & -3.40 & -5.40 & -6.55 & -7.15 \\
 A0 &  0.65 & -0.70 & -3.20 & -5.30 & -6.55 & -7.20 \\
 A2 &  1.30 & -0.40 & -2.90 & -5.20 & -6.65 & -7.40 \\
 A5 &  1.95 &  0.05 & -2.70 & -5.00 & -6.75 & -7.80 \\
 A8 &  2.40 &  0.40 & -2.50 & -4.85 & -6.75 & -8.15 \\
 F0 &  2.70 &  0.60 & -2.50 & -4.75 & -6.70 & -8.30 \\
 F2 &  3.00 &  0.80 & -2.50 & -4.70 & -6.60 & -8.30 \\
 F5 &  3.50 &  1.00 & -2.30 & -4.60 & -6.55 & -8.20 \\
 F8 &  4.00 &  1.00 & -2.30 & -4.55 & -6.45 & -8.00 \\
 G0 &  4.40 &  0.95 & -2.30 & -4.50 & -6.40 & -8.00 \\
 G2 &  4.70 &  0.90 & -2.30 & -4.50 & -6.30 & -8.00 \\
 G5 &  5.10 &  0.85 & -2.30 & -4.50 & -6.20 & -7.95 \\
 G8 &  5.50 &  0.75 & -2.30 & -4.45 & -6.15 & -7.85 \\
 K0 &  5.90 &  0.60 & -2.30 & -4.40 & -6.10 & -7.75 \\
 K2 &  6.30 &  0.30 & -2.30 & -4.40 & -6.00 & -7.70 \\
 K5 &  7.30 & -0.10 & -2.30 & -4.40 & -5.90 & -7.50 \\
 M0 &  8.80 & -0.40 & -2.40 & -4.60 & -5.70 & -7.10 \\
 M1 &  9.40 & -0.45 & -2.40 & -4.65 & -5.65 & -7.00 \\
 M2 &  10.1 & -0.45 & -2.40 & -4.75 & -5.60 & -6.95 \\
 M3 &  10.7 & -0.50 & -2.45 & -4.80 & -5.60 & -6.90 \\ \hline
\end{tabular}
\end{center}
\end{table}

\begin{figure}[!t]
\centering
\epsfxsize=8cm \epsfbox{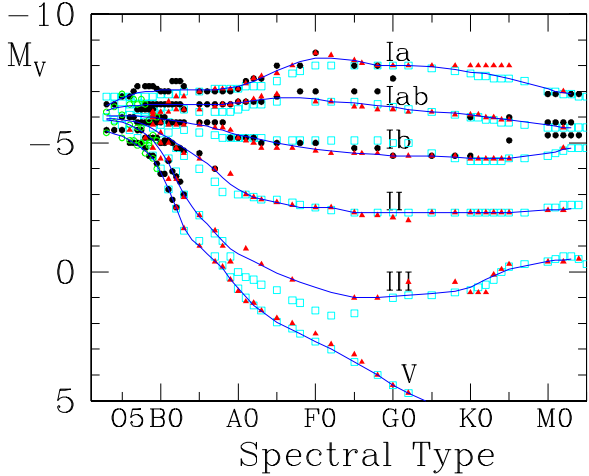} 
\caption{Spectral type - M$_V$ relation. Solid lines represent the adopted
         relations. Triangles, squares, dots, and open circles represent 
         the relations by \citet{blaauw63}, \citet{schmidt-kaler82}, 
         \citet{humphreys84}, and \citet{conti83}, respectively.
         \label{fig_sp_mv}}
\end{figure}

The adopted ZAMS relations for several colors are presented in Figure 
\ref{fig_zams}. The ZAMS relations by \citet{blaauw63},
\citet{mermilliod81}, and \citet{schmidt-kaler82} are compared in the figure.
In the lower panel, the ZAMS relation of the upper MS band
is compared with those of \citet{hanson97}, \citet{brott11}, and 
\citet{ekstrom12} which are transformed using the Sp - color relations
of Figure \ref{fig_sp_ci}. For \citet{brott11}, we took 10 models with the Milky
Way abundance and nearly the same initial surface velocity for a given mass
as those in Table 2 of \citet{ekstrom12}. The ZAMS of \citet{hanson97} is 
slightly brighter than the adopted ZAMS, while that of \citet{brott11}
and \citet{ekstrom12} is slightly fainter.

\subsection{Spectral Type - M$_V$ Relation}

One may determine $R_V$ from the [$V-M_V$, $E(B-V)$] diagram.
The absolute magnitude of a star can be determined from the ZAMS relation or
from the Sp - $M_V$ relation for a given LC. In addition, one has to decide the
membership of rare evolved stars in open clusters (see \citet{lim13} for
example). For these reasons
we have to adopt the Sp - $M_V$ relation. The usefulness of this
relation is very limited because of the uncertainty in the LC.
For O-type stars or evolved supergiant stars the scatter of $M_V$ is 
very large for a given Sp and LC because of their
rarity and intrinsic variety of their characteristics.

Although there are several uncertainties, we adopt the Sp - $M_V$ relation 
in order to have a constraint for the membership of supergiant stars and 
early-type stars. We present the adopted Sp - $M_V$ relation in Figure 
\ref{fig_sp_mv} and in Table \ref{tab_sp_mv}. The results from \citet{blaauw63},
\citet{schmidt-kaler82}, \citet{conti83}, and \citet{humphreys84} are compared
in the figure. The $M_V$ value of A -- F giants of \citet{schmidt-kaler82} is 
somewhat fainter than the values of \citet{blaauw63}, probably because of the
uncertainty of LC or the inclusion of subgiants, or both. The $M_V$ of 
\citet{humphreys84} is
slightly brighter for M-type supergiants (LC: Ib). We mainly adopt the Sp - 
$M_V$ relation of \citet{blaauw63} for late-type stars.

\subsection{Spectral Type - Temperature Relations}

\begin{figure}[!t]
\centering
\epsfxsize=8cm \epsfbox{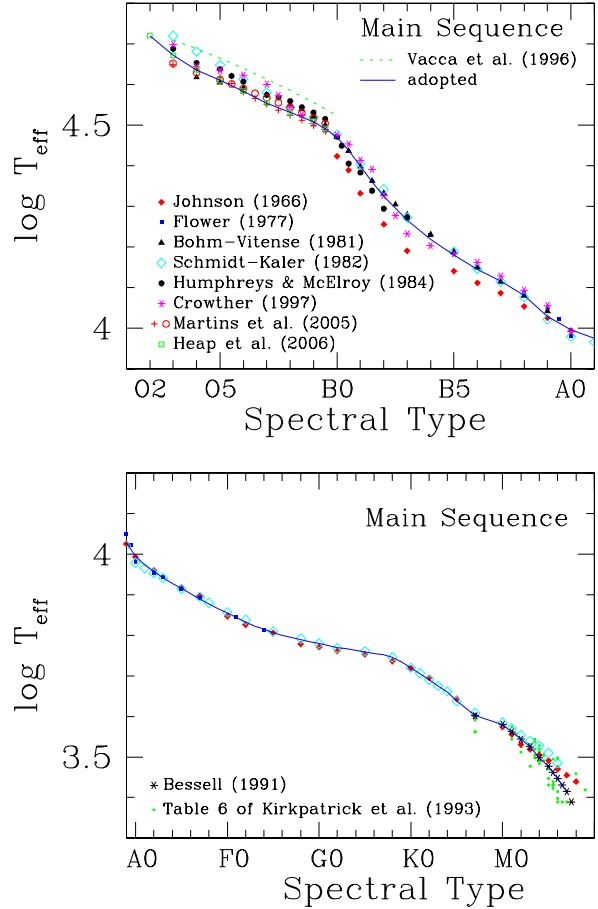} 
\caption{Spectral type-effective temperature relation for MS stars. Solid
         lines represent the adopted relations.
         \label{fig_sp_te_ms}}
\end{figure}

One important aim of the SOS is to test observationally the stellar 
evolution theory. There are two observational tests - one is to compare
observations with theoretical isochrones in the observational CMDs, 
the other is to compare the Hertzsprung-Russell diagrams (HRDs). In the first 
case, we have to use synthetic colors from model atmospheres, such as 
\citet{bessell98}. For the second case, we need to transform
observational quantities into physical parameters, such as T$_{\rm eff}$
and BC. Another important aim of the SOS is
to address whether the stellar IMF is universal or not. To derive the IMF of
an open cluster, one can directly determine the mass of the star in the HRD
or rely on the mass-luminosity relation from the theoretical isochrones.
As many young open clusters show a non-negligible spread in stellar age,
the use of theoretical mass-luminosity relations may be limited to the
intermediate-age or old open clusters. To construct the HRD of an open
cluster, we should adopt the relations between Sp and T$_{\rm eff}$, and between
T$_{\rm eff}$ and BC. In this section, we describe the relation between Sp and 
T$_{\rm eff}$ for a given LC.

\subsubsection{Main Sequence Stars}

The color - T$_{\rm eff}$ relation has been studied by several authors. As shown
in Figure \ref{fig_sp_te_ms}, the T$_{\rm eff}$ scales are well consistent with
each other for A -- K stars, but shows a large scatter for hot or cool stars.
For early-type stars, $B-V$ changes by only about 0.3 mag for O- and B-type stars
(T$_{\rm eff}$ = 10,000 -- 50,000 K) while $U-B$ changes by about 1.2 mag for 
the same T$_{\rm eff}$ range, but only shows a small change for O-type stars
\citep{massey85}. Therefore, although there is about 1 subclass uncertainty 
in spectral classification for O-type stars, Sp is the most important indicator
of T$_{\rm eff}$ for them. We first derive the Sp - T$_{\rm eff}$ 
relation for MS stars. T$_{\rm eff}$ for a given O-type is generally lower
for recent determinations than for older. For example, the T$_{\rm eff}$ scale 
of \citet{crowther97} of O-type stars is slightly lower than that of \citet{
vacca96}, but higher than that of \citet{martins05} or \citet{heap06}. We adopt
the recent T$_{\rm eff}$ scale of \citet{martins05} (their observational 
T$_{\rm eff}$ scale) and that of \citet{heap06} for O-type stars.

The T$_{\rm eff}$ scales of B-type stars, especially for early B-type stars show
a large scatter. The relation of \citet{johnson66} gives the lowest T$_
{\rm eff}$, while others indicate a slightly higher T$_{\rm eff}$. In addition, 
there is a large change in T$_{\rm eff}$ for early B-type stars. Unfortunately,
the T$_{\rm eff}$ scale of these stars did not attract recent studies. We adopt
the mean value of all the determinations except for \citet{johnson66} for
B-type stars.

The T$_{\rm eff}$ scales of M-type stars also show a large scatter. 
The T$_{\rm eff}$ scale of \citet{johnson66} or \citet{schmidt-kaler82} is
slightly higher, and that of \citet{bessell91} slightly lower. The T$_{\rm 
eff}$ of \citet{bessell91} is well-consistent with that of \citet{
kirkpatrick93}. We adopt the T$_{\rm eff}$ scale of \citet{bessell91} for 
M-type stars. For A -- K-type stars, we take the mean value of \citet{
johnson66}, \citet{flower77}, and \citet{schmidt-kaler82}. The adopted 
T$_{\rm eff}$ scale for MS stars is shown with a solid line in Figure 
\ref{fig_sp_te_ms}, and reported in Table \ref{tab_sp_te_ci}.

\begin{figure}[!t]
\centering
\epsfxsize=8cm \epsfbox{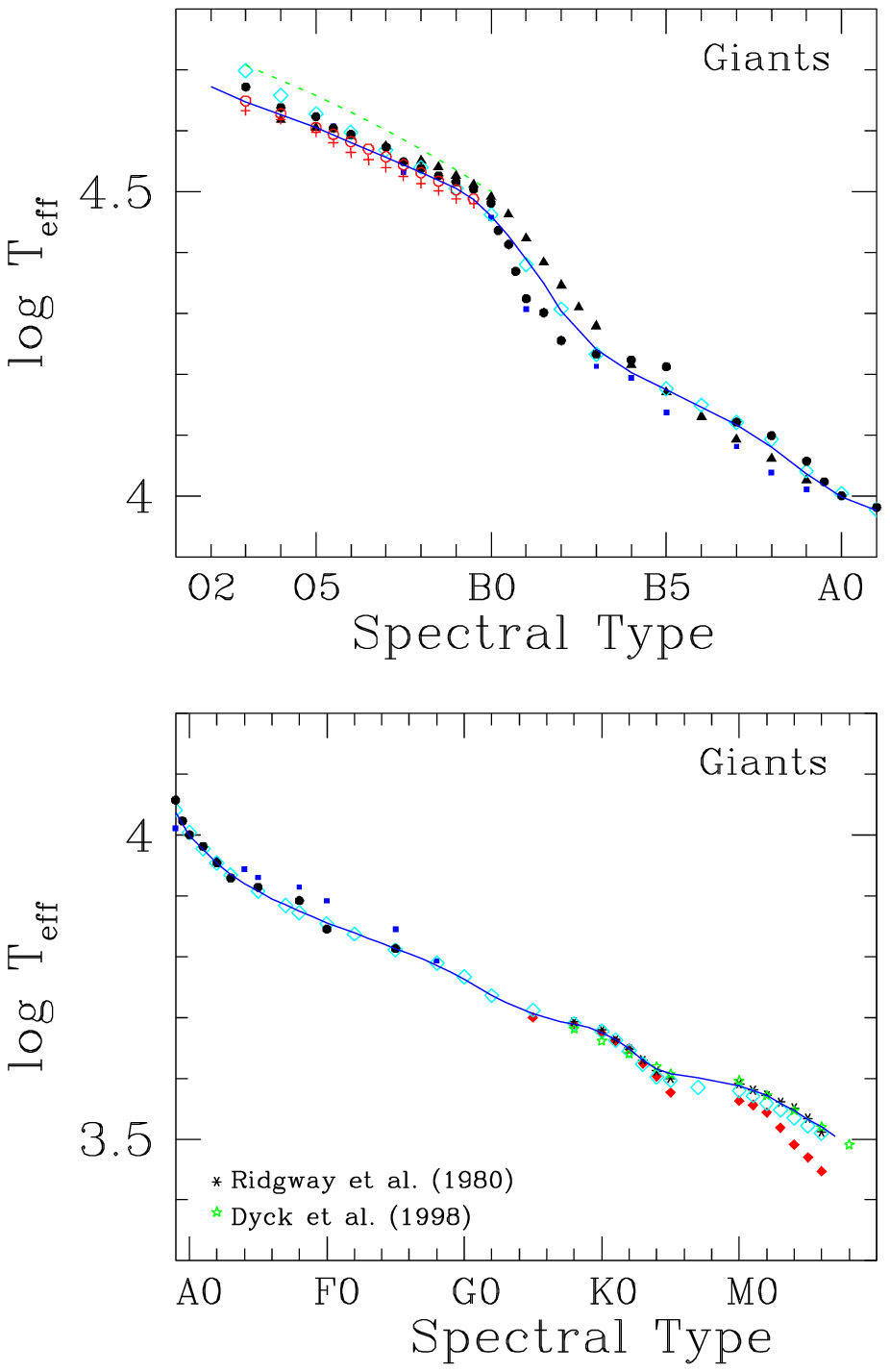} 
\caption{Spectral-type-effective temperature relation for giant stars. 
         Solid lines represent the adopted relations. Symbols are
         the same as in Figure \ref{fig_sp_te_ms} except for those in 
         the lower panel. \label{fig_sp_te_gt}}
\end{figure}

\subsubsection{Giant Stars}

The T$_{\rm eff}$ scales of O-type giants show the same trend - lower  
T$_{\rm eff}$ for recent determinations. As for the case of MS stars, we choose 
the T$_{\rm eff}$ scale of \citet{martins05}
for O-type stars. For B-type stars, the T$_{\rm eff}$ scale
of \citet{schmidt-kaler82} is intermediate between \citet{bohm-vitense81} and
\citet{humphreys84} or \citet{flower77}. We adopt the T$_{\rm eff}$ scale of
\citet{schmidt-kaler82} for B-type stars.

For late-K and M-type stars, the T$_{\rm eff}$ scale of \citet{johnson66} is
lower than the others. We take the T$_{\rm eff}$ scale of \citet{ridgway80}
for K and M-type stars. For A -- K-type stars, \citet{flower77} gives slightly
higher T$_{\rm eff}$. We mainly take \citet{schmidt-kaler82} for A -- K-type
stars. The adopted Sp - T$_{\rm eff}$ relation for giants is presented in
Figure \ref{fig_sp_te_gt} and in Table \ref{tab_sp_te_ci}.

\begin{figure}[!t]
\centering
\epsfxsize=8cm \epsfbox{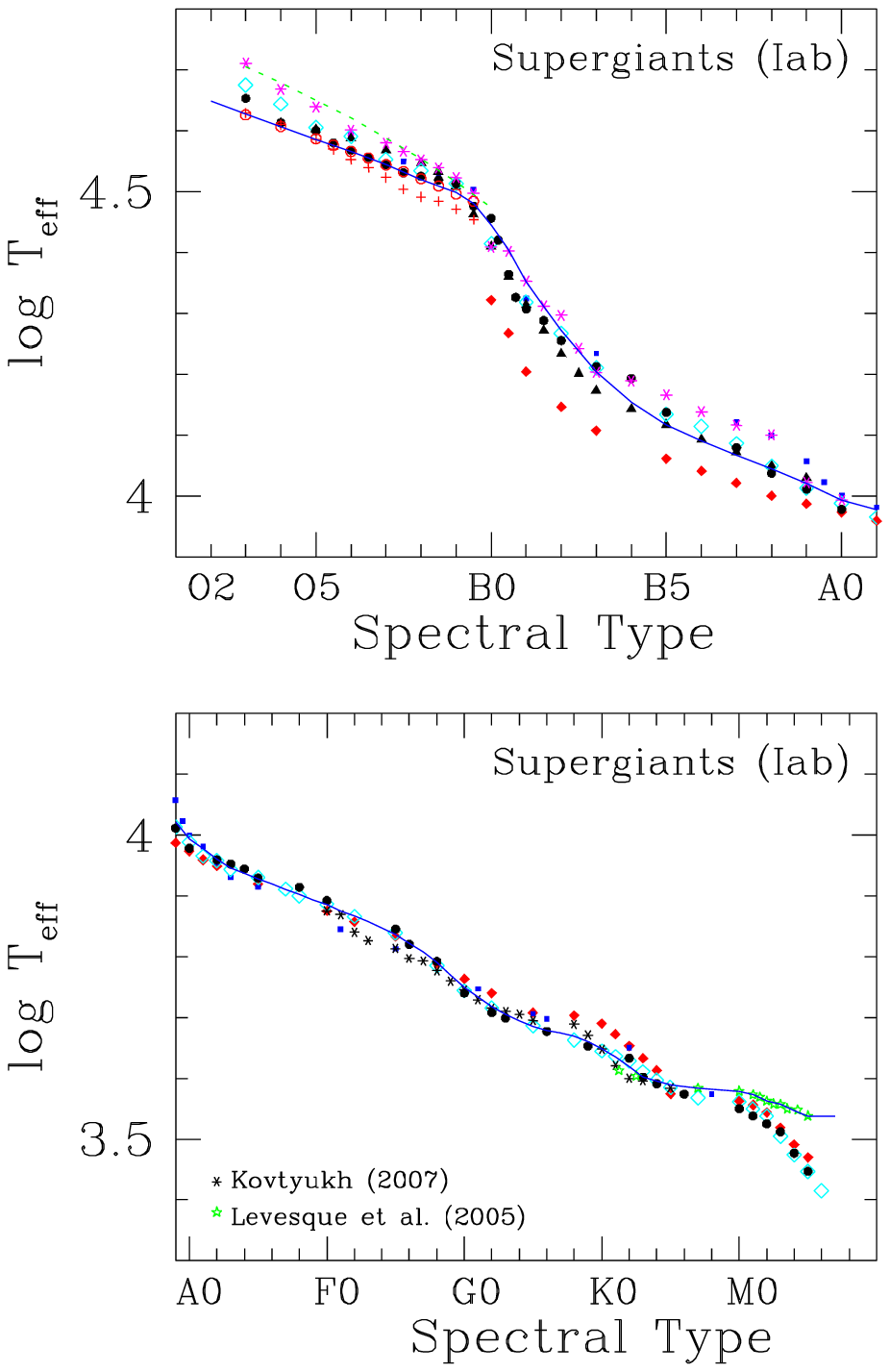} 
\caption{Spectral type-effective temperature relation for supergiant stars. 
         Solid lines represent the adopted relations. Symbols are 
         the same as in Figure \ref{fig_sp_te_ms} except for those in 
         the lower panel. \label{fig_sp_te_sg}}
\end{figure}

\subsubsection{Supergiant Stars}

The T$_{\rm eff}$ scale of early-type stars shows a large scatter as presented
in Figure \ref{fig_sp_te_sg}. The T$_{\rm eff}$ of B-type stars of \citet{
johnson66} is lower than the others. As is in the previous sections, we take
the T$_{\rm eff}$ scale of \citet{martins05} for O-type stars, that of 
\citet{crowther97} for early B stars, and that of \citet{bohm-vitense81}
or \citet{schmidt-kaler82} for late B stars.

For A -- G-type stars, the T$_{\rm eff}$ of \citet{humphreys84} is well
consistent with that of \citet{schmidt-kaler82}. We take the average of their
relations. Recently \citet{levesque05} showed that there is a lower limit
in the T$_{\rm eff}$ of M-type supergiants, and their T$_{\rm eff}$ scale of
M-type is much higher than the others. We adopt the T$_{\rm eff}$ scale of
\citet{levesque05} for M-type supergiants. The adopted Sp - T$_{\rm eff}$ 
relation is shown in Figure \ref{fig_sp_te_sg} and reported in
Table \ref{tab_sp_te_ci}.

\subsection{Spectral Type - Color and Color - Temperature Relations}

\begin{figure}[!t]
\centering
\epsfxsize=7.2cm \epsfbox{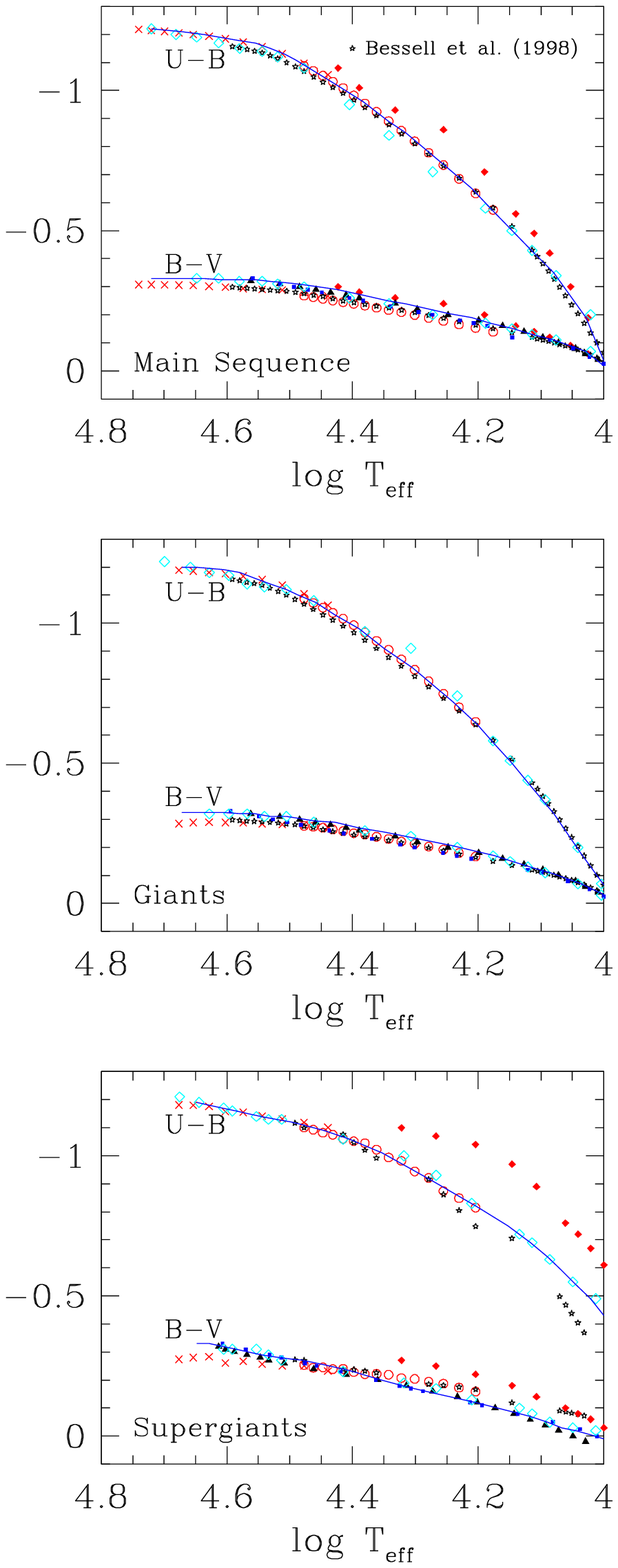} 
\caption{Color - temperature relations for OB stars. 
	 Open circles, crosses, and star symbols represent the synthetic color
 	 from Tlusty O-type star models, Tlusty B-type star models, and 
	 \citet{bessell98} for MS stars (upper panel), for giant stars (middle 
         panel), and for supergiant stars (LC = Iab, lower panel), respectively.
	 The T$_{\rm eff}$ - $\log ~g$ relation for a given LC is taken from
	 \citet{sung95}. The other symbols are the same as those in Figure 
	 \ref{fig_sp_te_ms}. Solid lines represent the adopted relations.
         \label{fig_col_te}}
\end{figure}

\begin{figure*}[!t]
\centering
\epsfxsize=16cm \epsfbox{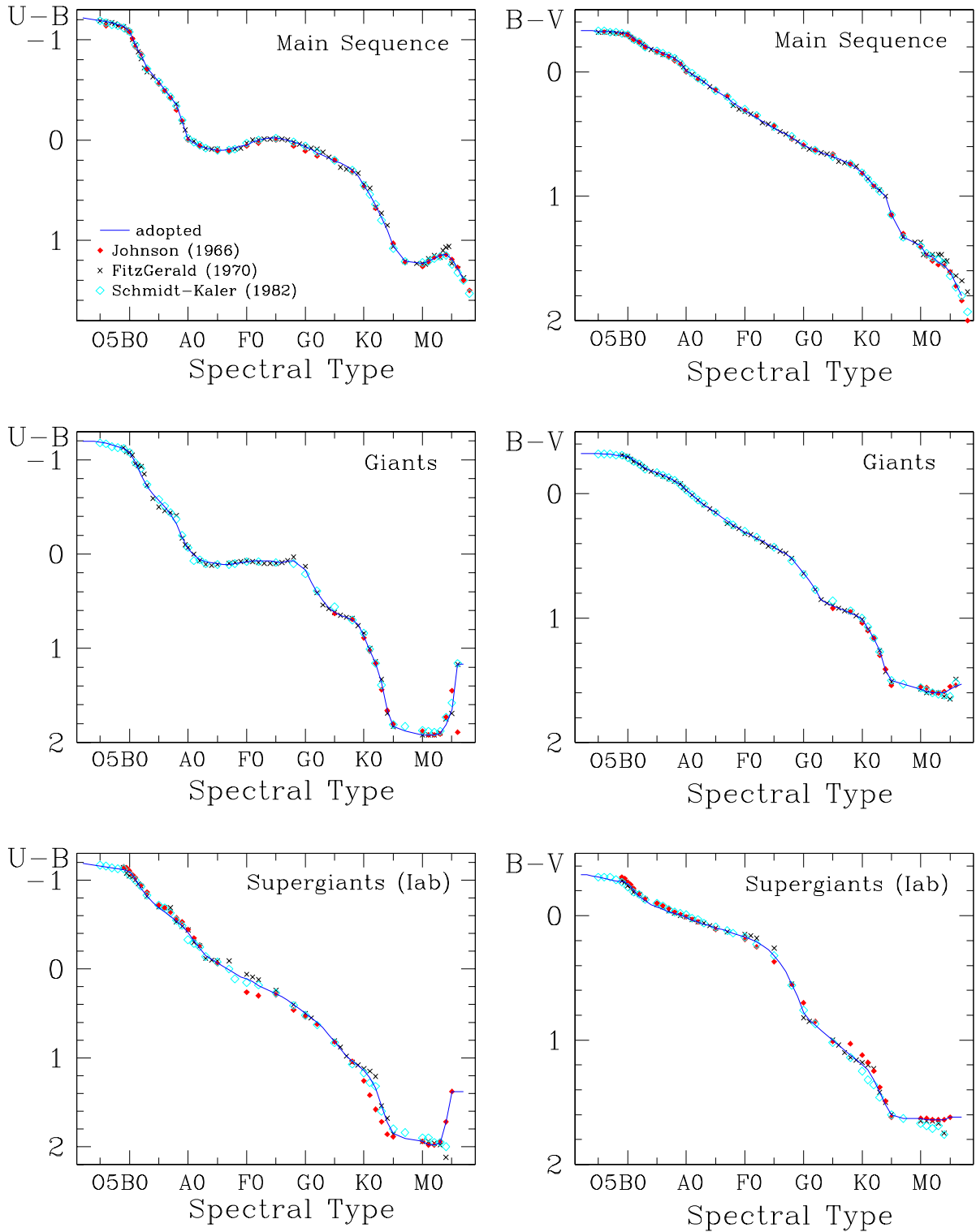} 
\caption{Spectral type-color relations.
         Solid lines represent the adopted relations.
         \label{fig_sp_ci}}
\end{figure*}

There are several studies on the Sp - color relations. Among them,
\citet{fitzgerald70} analyzed extensive data in the $UBV$ photoelectric 
catalogue of \citet{blanco68}. In general, these results are well-consistent 
with each other, as shown in Figure \ref{fig_sp_ci}. However, in some cases,
such as for B- and K-type stars where $U-B$ changes rapidly, it is not easy to
take the intrinsic color for a given Sp. For such cases, we inevitably use
the color-color relations adopted in Section 4.1. In addition, as the intrinsic
color of early-type stars is still uncertain, we use the T$_{\rm eff}$ - color 
relations for early-type stars subsidarily (see Figure \ref{fig_col_te}).

The $U-B$ colors from Tlusty models are in general well-consistent with the 
empirical relations, but the $B-V$ colors are slightly redder than the adopted
relation by about 0.02 mag for MS and giants. The same is true for 
\citet{bessell98}. In addition, the $U-B$ of \citet{bessell98} is slightly
redder than the synthetic colors from Tlusty models for O- and early B-type
stars. The color - T$_{\rm eff}$ relation of supergiant stars shows a large
scatter. Nevertheless, the synthetic $U-B$ colors from Tlusty models are
well-consistent with the $U-B$ versus T$_{\rm eff}$ relation of \citet{
schmidt-kaler82}. We adopt the $B-V$ versus T$_{\rm eff}$ relation using
the color - color relation in Figure \ref{fig_cc} for supergiant stars (Iab). 
The adopted relation is well-consistent with the empirical relation by 
\citet{bohm-vitense81} or \citet{schmidt-kaler82}, but shows a different trend
relative to the synthetic $B-V$ from the Tlusty models. The $U-B$ of 
\citet{bessell98} shows a large scatter due to the current limitation
in models with $\log~ g$ appropriate to supergiant stars (see \citet{sung95}
for the T$_{\rm eff}$ versus $\log~ g$ relation). The adopted Sp - color
relations are presented in Figure \ref{fig_sp_ci} and in Table
\ref{tab_sp_te_ci}.

\subsection{Temperature - Bolometric Correction}

\begin{table*}[!t]
\begin{center}
\centering
\caption{Spectral type-T$_{\rm eff}$-color relation \label{tab_sp_te_ci} }
\doublerulesep2.0pt
\renewcommand\arraystretch{1.0}
\begin{tabular}{c||r@{}r@{}rr|r@{}r@{}rr|r@{}r@{}rr}
\hline \hline
L.C. & \multicolumn{4}{c}{V} & \multicolumn{4}{c}{III} & \multicolumn{4}{c}{Iab} \\
Sp. Type& T$_{\rm eff}$ & $B-V$ & $U-B$ & B.C. & T$_{\rm eff}$ & $B-V$ & $U-B$ & B.C. & T$_{\rm eff}$ & $B-V$ & $U-B$ & B.C. \\ \hline
 O2 &  4.720 & -0.33 & -1.22 & -4.52 &  4.672 & -.325 & -1.20 & -4.22 &  4.648 & -0.33 & -1.19 & -4.06 \\
 O3 &  4.672 & -0.33 & -1.21 & -4.19 &  4.647 & -.325 & -1.20 & -4.02 &  4.628 & -0.33 & -1.18 & -3.90 \\
 O4 &  4.636 & -0.33 & -1.20 & -3.94 &  4.626 & -.325 & -1.19 & -3.88 &  4.607 & -0.32 & -1.17 & -3.73 \\
 O5 &  4.610 & -.325 & -1.19 & -3.77 &  4.605 & -.325 & -1.19 & -3.72 &  4.585 & -0.31 & -1.16 & -3.57 \\
 O6 &  4.583 & -.325 & -1.18 & -3.58 &  4.580 & -0.32 & -1.18 & -3.54 &  4.565 & -0.30 & -1.15 & -3.43 \\
 O7 &  4.554 & -.325 & -1.17 & -3.39 &  4.556 & -0.32 & -1.16 & -3.39 &  4.544 & -0.29 & -1.14 & -3.29 \\
 O8 &  4.531 & -0.32 & -1.15 & -3.23 &  4.531 & -0.31 & -1.14 & -3.21 &  4.519 & -0.28 & -1.13 & -3.11 \\
 O9 &  4.508 & -.315 & -1.13 & -3.03 &  4.505 & -0.31 & -1.12 & -3.05 &  4.498 & -.275 & -1.12 & -2.96 \\
 B0 &  4.470 & -.305 & -1.08 & -2.84 &  4.459 & -.295 & -1.07 & -2.73 &  4.445 & -.255 & -1.09 & -2.66 \\
 B1 &  4.400 & -.275 & -0.98 & -2.40 &  4.389 & -0.27 & -0.98 & -2.41 &  4.353 & -0.20 & -1.01 & -2.13 \\
 B2 &  4.325 & -0.24 & -0.87 & -2.02 &  4.304 & -.235 & -0.84 & -1.95 &  4.272 & -.155 & -0.91 & -1.70 \\
 B3 &  4.265 & -0.21 & -0.75 & -1.62 &  4.240 & -.205 & -0.72 & -1.56 &  4.203 & -0.12 & -0.82 & -1.31 \\
 B5 &  4.180 & -0.17 & -0.58 & -1.22 &  4.175 & -0.17 & -0.57 & -1.19 &  4.117 & -0.07 & -0.69 & -0.87 \\
 B6 &  4.145 & -0.15 & -0.50 & -1.02 &  4.146 & -0.15 & -0.50 & -1.02 &  4.090 & -0.05 & -0.64 & -0.73 \\
 B7 &  4.115 & -0.13 & -0.43 & -0.85 &  4.117 & -0.13 & -0.42 & -0.86 &  4.067 & -0.03 & -0.59 & -0.60 \\
 B8 &  4.080 & -0.11 & -0.35 & -0.66 &  4.080 & -.105 & -0.32 & -0.66 &  4.044 & -0.02 & -0.54 & -0.49 \\
 B9 &  4.028 & -0.07 & -0.19 & -0.39 &  4.037 & -0.07 & -0.18 & -0.44 &  4.021 & -.005 & -0.49 & -0.38 \\
 A0 &  3.995 & -0.01 & -0.01 & -0.24 &  3.998 & -0.03 & -0.06 & -0.25 &  3.993 &  .015 & -0.41 & -0.26 \\
 A1 &  3.974 &  0.02 &  0.03 & -0.15 &  3.976 &  0.01 &  0.01 & -0.15 &  3.977 &  .035 & -0.32 & -0.20 \\
 A2 &  3.958 &  0.05 &  0.06 & -0.08 &  3.954 &  0.05 &  0.06 & -0.07 &  3.960 &  0.05 & -0.23 & -0.12 \\
 A3 &  3.942 &  0.08 &  0.08 & -0.03 &  3.935 &  0.09 &  0.08 &  0.00 &  3.946 &  0.07 & -0.15 & -0.05 \\
 A5 &  3.915 &  0.15 &  0.10 &  0.00 &  3.907 &  0.15 &  0.10 &  0.05 &  3.928 &  0.10 & -0.07 &  0.00 \\
 A6 &  3.902 &  0.18 &  0.10 &  0.01 &  3.895 &  0.19 &  0.11 &  0.06 &  3.920 &  0.11 & -0.03 &  0.03 \\
 A7 &  3.889 &  0.21 &  0.09 &  0.02 &  3.885 &  0.22 &  0.11 &  0.06 &  3.910 &  .125 &  0.01 &  0.05 \\
 A8 &  3.877 &  0.25 &  0.08 &  0.02 &  3.875 &  0.25 &  0.10 &  0.06 &  3.902 &  0.14 &  0.05 &  0.07 \\
 F0 &  3.855 &  0.31 &  0.05 &  0.01 &  3.855 &  0.31 &  0.08 &  0.05 &  3.885 &  0.17 &  0.11 &  0.10 \\
 F1 &  3.843 &  0.34 &  0.02 &  0.01 &  3.847 &  0.33 &  0.07 &  0.04 &  3.875 &  0.19 &  0.14 &  0.11 \\
 F2 &  3.832 &  0.37 &  0.00 &  0.00 &  3.839 &  0.36 &  0.07 &  0.04 &  3.867 &  0.21 &  0.19 &  0.12 \\
 F3 &  3.822 &  0.40 & -0.01 &  0.00 &  3.830 &  0.38 &  0.07 &  0.03 &  3.858 &  0.24 &  0.22 &  0.12 \\
 F5 &  3.806 &  0.45 & -0.02 & -0.01 &  3.813 &  0.43 &  0.08 &  0.02 &  3.836 &  0.32 &  0.28 &  0.11 \\
 F6 &  3.800 &  0.48 & -0.01 & -0.02 &  3.805 &  0.46 &  0.09 &  0.01 &  3.822 &  0.38 &  0.31 &  0.10 \\
 F7 &  3.794 &  0.50 &  0.00 & -0.02 &  3.796 &  0.48 &  0.07 &  0.00 &  3.807 &  0.45 &  0.35 &  0.08 \\
 F8 &  3.789 &  0.53 &  0.02 & -0.03 &  3.785 &  0.52 &  0.07 & -0.01 &  3.790 &  0.55 &  0.40 &  0.05 \\
 G0 &  3.780 &  0.59 &  0.07 & -0.04 &  3.763 &  0.64 &  0.17 & -0.05 &  3.750 &  0.78 &  0.50 & -0.05 \\
 G1 &  3.775 &  0.61 &  0.09 & -0.04 &  3.750 &  0.70 &  0.30 & -0.08 &  3.734 &  0.84 &  0.55 & -0.09 \\
 G2 &  3.770 &  0.63 &  0.13 & -0.05 &  3.737 &  0.77 &  0.41 & -0.11 &  3.718 &  0.88 &  0.60 & -0.16 \\
 G3 &  3.767 &  0.65 &  0.15 & -0.06 &  3.725 &  0.85 &  0.49 & -0.15 &  3.705 &  0.92 &  0.66 & -0.21 \\
 G5 &  3.759 &  0.68 &  0.21 & -0.07 &  3.706 &  0.90 &  0.62 & -0.22 &  3.685 &  1.00 &  0.82 & -0.32 \\
 G6 &  3.755 &  0.70 &  0.23 & -0.08 &  3.700 &  0.92 &  0.65 & -0.25 &  3.679 &  1.04 &  0.90 & -0.35 \\
 G7 &  3.752 &  0.72 &  0.26 & -0.09 &  3.693 &  0.94 &  0.67 & -0.28 &  3.675 &  1.08 &  0.98 & -0.37 \\
 G8 &  3.745 &  0.74 &  0.30 & -0.10 &  3.689 &  0.96 &  0.70 & -0.30 &  3.670 &  1.12 &  1.04 & -0.40 \\
 K0 &  3.720 &  0.81 &  0.45 & -0.18 &  3.675 &  1.01 &  0.87 & -0.37 &  3.648 &  1.20 &  1.14 & -0.54 \\
 K1 &  3.705 &  0.86 &  0.54 & -0.24 &  3.664 &  1.08 &  1.02 & -0.44 &  3.635 &  1.24 &  1.22 & -0.65 \\
 K2 &  3.690 &  0.91 &  0.65 & -0.32 &  3.648 &  1.16 &  1.16 & -0.54 &  3.619 &  1.32 &  1.34 & -0.80 \\
 K3 &  3.675 &  0.96 &  0.77 & -0.41 &  3.630 &  1.26 &  1.36 & -0.68 &  3.602 &  1.42 &  1.54 & -0.98 \\
 K5 &  3.638 &  1.15 &  1.06 & -0.65 &  3.607 &  1.50 &  1.83 & -0.92 &  3.589 &  1.60 &  1.85 & -1.14 \\
 M0 &  3.580 &  1.40 &  1.23 & -1.18 &  3.588 &  1.57 &  1.92 & -1.14 &  3.579 &  1.63 &  1.94 & -1.27 \\
 M1 &  3.562 &  1.47 &  1.21 & -1.39 &  3.580 &  1.59 &  1.92 & -1.25 &  3.573 &  1.64 &  1.98 & -1.35 \\
 M2 &  3.544 &  1.49 &  1.18 & -1.64 &  3.572 &  1.60 &  1.92 & -1.37 &  3.563 &  1.64 &  1.98 & -1.53 \\
 M3 &  3.525 &  1.53 &  1.15 & -2.02 &  3.559 &  1.60 &  1.90 & -1.64 &  3.557 &  1.64 &  1.96 & -1.64 \\
 M4 &  3.498 &  1.56 &  1.14 & -2.55 &  3.547 &  1.60 &  1.81 & -1.90 &  3.547 &  1.64 &  1.72 & -1.82 \\
 M5 &  3.477 &  1.61 &  1.19 & -3.05 &  3.533 &  1.57 &  1.65 & -2.22 &  3.538 &  1.62 &  1.38 & -2.05 \\
\hline
\end{tabular}
\end{center}
\end{table*}

\begin{figure}[!t]
\centering
\epsfxsize=7.5cm \epsfbox{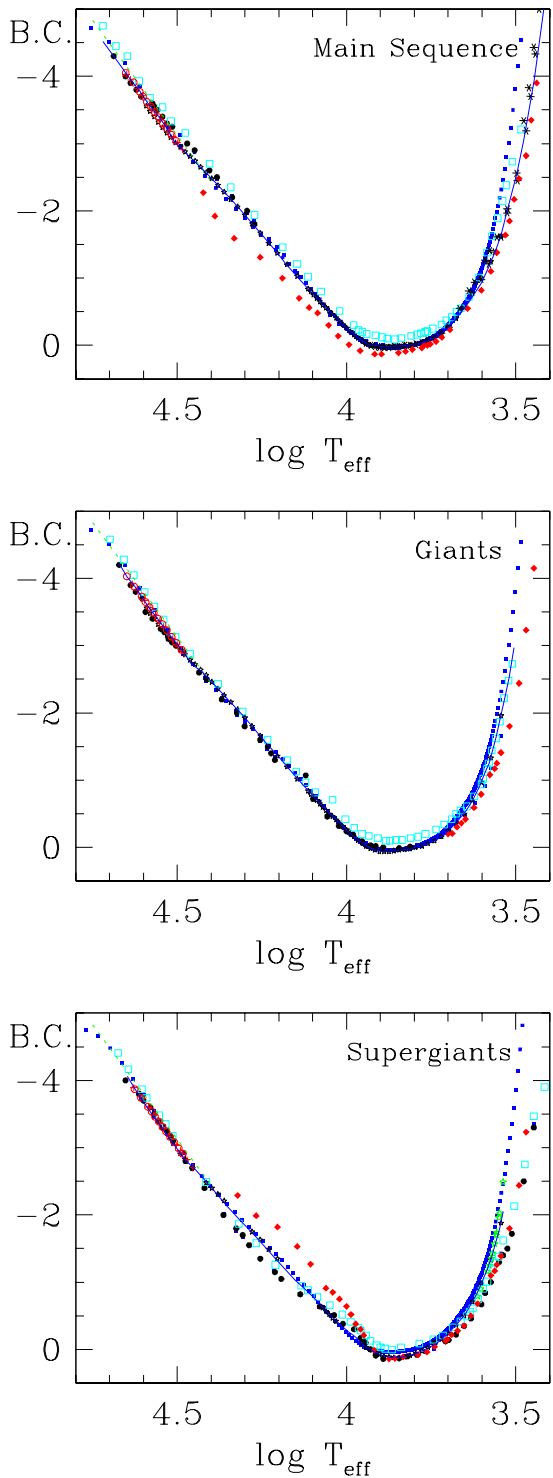} 
\caption{Temperature-bolometric correction relations. 
         Solid lines represent the adopted relations.
         \label{fig_te_bc}}
\end{figure}

To construct the HRD, we need to correct for the amount of radiation emitted in
the ultraviolet and in the infrared pass band. \citet{johnson66} assumed BC = 
0.00 for the Sun, and presented relatively smaller BC as shown in Figure
\ref{fig_te_bc}. On the other hand, \citet{schmidt-kaler82} adopted 
BC = -0.19 for the Sun based on the \citet{kurucz79} model atmospheres,
and published relatively larger BC for all cases. Most astronomers now adopt 
$M_{bol}$ = 4.75 and BC = -0.07 for the Sun. \citet{bessell98} discussed
this issue in detail.

As can be seen in Figure \ref{fig_te_bc}, BC is a function of 
T$_{\rm eff}$ for a given LC. \citet{vacca96} mentioned that BC for O-type
stars is essentially a function of T$_{\rm eff}$ only, i.e. independent of 
surface gravity. \citet{martins05} presented a slightly smaller correction than
that of \citet{vacca96}, but BC is independent of surface gravity for O-type 
stars. \citet{balona94} presented the relation between 
T$_{\rm eff}$ and BC for from O9 to G5 stars. \citet{flower96} gave
the same BC scale for MS and giant stars. \citet{bessell98} also calculated
BC for a model atmosphere with various surface gravities, but there are 
non-negligible differences in BC among different LCs. 

We assume that BC for O stars is independent of surface gravity, but that for 
B -- M stars differs for different LCs. For M stars, we adopt the BC scale of
\citet{bessell91} for MS stars, \citet{schmidt-kaler82} for giant stars, and
\citet{levesque05} for supergiant stars. For B -- K stars, we adopt the BC
scale of \citet{bessell98}. The adopted BC scales are presented in Figure
\ref{fig_te_bc} and in Table \ref{tab_sp_te_ci}.

\section{DATA ANALYSIS TOOLS}

\subsection{Reddening Law}

To determine the physical parameters of stars and clusters accurately, we should
know the correct interstellar reddening. Many investigators determine
the total extinction $A_V$ by assuming the total-to-selective extinction
ratio $R_V$ to be 3.1. However, the interstellar reddening law $R_V$ is
known to be different for different sightlines \citep{fitzpatrick09}. 
In addition, many young open clusters are known to show an abnormal reddening 
law. Recently Sung et al. (2013, in preparation) confirmed the variation of the 
reddening law with Galactic longitude from an analysis of optical and 
2MASS data for about 200 young open clusters.

\begin{figure}[!t]
\centering
\epsfxsize=8cm \epsfbox{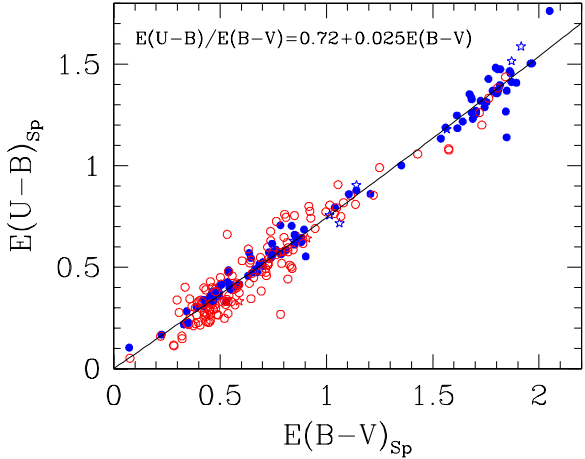} 
\caption{Color excess ratio between $E(B-V)_{sp}$ and $E(U-B)_{sp}$ for
	 O- and early B- (Sp $\leq$ B2.5) type stars in young open clusters.
	 Dots and circles represent O- and B-type MS stars, respectively, while
         star symbols denote evolved stars. 
 	 \label{fig_exr}}
\end{figure}

Before determining the color excess ratio,
we should determine the reddening $E(B-V)$ from the ($U-B,~ B-V$) TCD.
The slope of the reddening vector in the TCD is known to depend both
on the amount of reddening and on the intrinsic color (see \citet{golay74}).
We determined $E(B-V)$ and $E(U-B)$ for about 255 OB-stars in the young open
clusters (NGC 6530, $\eta$ Car, NGC 6611, NGC 6231, NGC 6823, IC 1805,
Westerlund 2, NGC 2244, NGC 2264, and the ONC) using the relation between
spectral type and intrinsic color (see Section 4.5), and this is shown in
Figure \ref{fig_exr}. From the figure, we could not find any difference
between O-type stars and B-type stars, which implies that the relation
between $E(B-V)$ and $E(U-B)$ depends only on the amount of reddening.
We adopt the excess ratio as follow;

\begin{equation}
E(U-B)/E(B-V) = 0.72 + 0.025 E(B-V).
\end{equation}

\begin{figure}[!b]
\centering
\epsfxsize=8cm \epsfbox{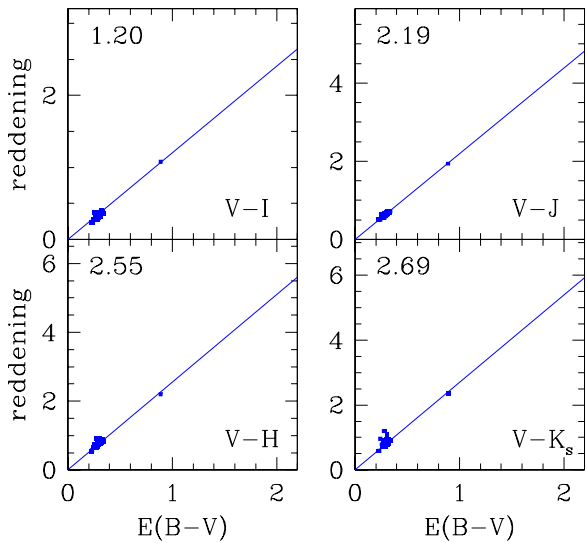} 
\caption{Color excess ratios for NGC 6531. The color excess of each color 
         is calculated using the relation between intrinsic colors (see 
         Figure \ref{fig_nir} and Table \ref{tab_ir}). The $UBVI$ CCD data are
         from \citet{park01}. All four color excess ratios imply that $R_V$
         = 2.96 $\pm$ 0.03 for NGC 6531. \label{fig_red} }
\end{figure}

\noindent
\citet{cardelli89} showed that the total extinction $A_\lambda$ could be
described as a simple function of $R_V$. If we assume their expression,
the color excess can also be expressed as $E(U-B)/E(B-V) = 0.859-0.045 R_V$.
If so, the ratio should be much smaller for young open clusters with 
an abnormal reddening law, e.g. the young open clusters in the $\eta$ Car nebula
\citep{hur12}.\footnote{The $U-B$ color of Herschel 36 in the young open 
cluster NGC 6530 is bluer than the $U-B$ expected from the color excess ratio.
However, the bluer $U-B$ of the star should be checked since it may be
due to the error in the photometry \citep{johnson67,sung00a} or to other 
effects, such as accretion.} We could not find any difference in color excess
ratios among O-type stars in the $\eta$ Car nebula, and therefore we neglect
the effect of $R_V$ on $E(U-B)/E(B-V)$.

Knowledge of the reddening law, especially the total-to-selective
extinction ratio $R_V \equiv A_V / E(B-V)$, is very important in estimating
the distance to the object. In general, $R_V$ can be determined from the color
excess ratio \citep{guetter89}. We will determine the $R_V$ of target clusters
using their relations,

\begin{equation}
R_V = 2.45 E(V-I)/E(B-V)
\end{equation}
\begin{equation}
 ~= 1.33 E(V-J)/E(B-V)
\end{equation}
\begin{equation}
 ~ = 1.17 E(V-H)/E(B-V)
\end{equation}
\begin{equation}
 ~= 1.10 E(V-K_s)/E(B-V).
\end{equation}

The reddening $E(B-V)$ of individual early-type stars is calculated from 
the ($U-B, ~B-V$) TCD, and the color excess of each color is calculated 
using the relation between intrinsic colors described in Section 4.1.
Figure \ref{fig_red} shows that the $R_V$ of NGC 6531 is 2.96 $\pm$ 0.03 from
all the four colors.

\subsection{Membership Selection Criteria}

As open clusters are in the Galactic plane, we can expect many field
interlopers in the foreground and background. Therefore, membership
selection is crucial in the studies of a cluster.
There are several different membership criteria
for different clusters or for different age groups. 

Proper motion studies are a classical membership criterion for cluster studies.
To obtain reliable proper motions for the stars in the cluster field,
a long baseline in time is very important. As old photographic plates
have very low sensitivity, their limiting magnitude is about $V$ = 15 mag,
which is far shallower than the limiting magnitude obtainable from a small
telescope and from a modern CCD camera. The currently available
CCD-based proper motion catalog from the US Naval Observatory, UCAC-3 \citep{
zacharias10} is in most cases useless for the membership selection of open
clusters at 1 kpc or farther, because there is nearly no difference in proper
motion between cluster members and field stars. The 10 $\mu as$
astrometric data from the {\it Gaia} astrometric satellite \citep{lindegren07,
turon12} will revolutionize the study of stellar astrophysics in the 2020s.

\subsubsection{Early-Type Stars}

Early-type members (Sp $\lesssim$ B5) in young or intermediate-age open clusters can be selected
from the ($U-B$, $B-V$) TCD without any ambiguity. Unfortunately there is nearly
no membership criterion for late-B or A -- F-type stars in reddened 
clusters. In such cases we have to estimate the number of members 
in a statistical way, but it is impossible to assign a membership individually.

\subsubsection{Young Open Clusters}

Several membership criteria are used to select the low-mass PMS
stars in young open clusters.
\citet{sung97} introduced H$\alpha$ photometry as a membership criterion
for low-mass PMS stars in the young open cluster NGC 2264, and successfully
selected many PMS members in the T Tauri stage. H$\alpha$ photometry
as a membership criterion is restricted to the extremely young open 
clusters with ages younger than about 5 Myr. Another method of membership
selection in the optical pass bands is to study the variability because most
young stars show variability due to mass accretion.

An important characteristic of young PMS stars is an IR excess
emission from their circumstellar disks. Unfortunately most PMS stars do not 
show any appreciable emission in the near-IR pass bands. Therefore,
the usefulness of near-IR $JHK$ photometry is very limited. An IR excess in
the mid-IR pass bands, such as the {\it Spitzer} IRAC bands, is an important
membership criterion for extreme young open clusters \citep{sung09}.
The probability of membership selection from a mid-IR excess
is very similar to that from H$\alpha$ photometry in the optical, but mid-IR 
excess is very useful for highly reddened embedded clusters.

While only classical T Tauri stars show very strong H$\alpha$ emission and an
IR excess, both classical and weak-line T Tauri stars are very bright in X-rays.
Therefore, X-ray emission is the most important membership criterion
for low-mass PMS stars in young open clusters. Only a few PMS stars with edge-on
disks do not show any appreciable emission or lack of emission \citep{sung04b}.
One weak point is that X-ray activity prolongs for a long time \citep{ sung08b}.
Hence, we need to pay attention to remove field interlopers with strong 
X-ray emission.

\subsubsection{Intermediate-Age and Old Open Clusters}

There is no reliable membership criterion for intermediate-age or old open
clusters, except the CMD. For nearby clusters, \citet{sung99a} devised
a photometric membership criterion using the merit of multicolor photometry.
Most intermediate-age or old open clusters do not show any appreciable
amount of differential reddening across the cluster. In addition, the effect 
of reddening differs for different colors. The distance modulus from the
($V,~ B-V$) CMD should be the same value [($V_0 - M_V)_{cl}$] as that 
from the ($V,~ V-I$) CMD if the star is a member of the cluster. 
As photometric errors, binarity and other effects such as chromospheric activity
\citep{sung02} or metallicity difference \citep{sung99a} can also affect 
the ZAMS relation the criterion for membership selection should be relaxed -
(i) the average value of the distance modulus should be in the
range [($V_0 - M_V)_{cl} - 0.75 - 2 \sigma_{V_0 - M_V}$] and [($V_0 - 
M_V)_{cl} + 2 \sigma_{V_0 - M_V}$], and (ii) the difference in the distance moduli
should be smaller than $2.5 \sigma_{V_0 - M_V}$.

For open clusters at 1 kpc or farther, we can expect many field interlopers
with similar photometric characteristics. For such cases we should statistically
estimate the number of field interlopers in the field region and subtract
them \citep{kook10}.

\subsection{Distance}

\begin{figure*}[!t]
\centering
\epsfxsize=15cm \epsfbox{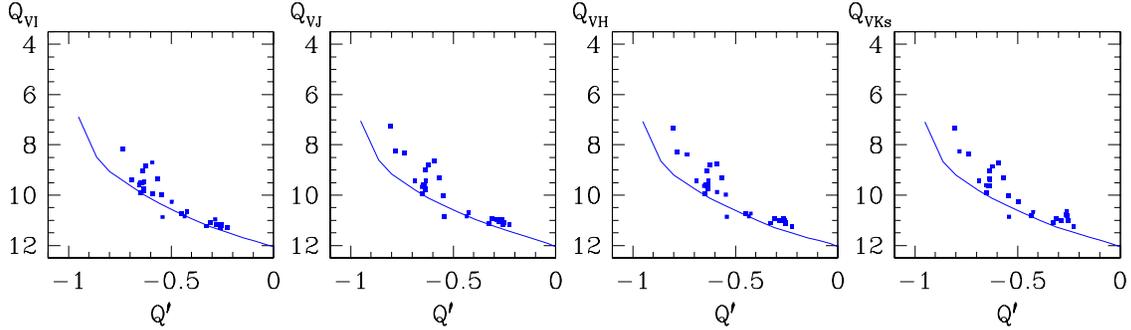} 
\caption{The reddening-independent color-magnitude diagrams of NGC 6531
         \citep{park01}. The distance modulus of 10.5 mag is applied for all
         CMDs. See the main text for details. \label{fig_dm}}
\end{figure*}

In general the distance to an open cluster is estimated using the ZAMS relation
in the reddening-corrected CMDs. Before 1990, most observations were performed
in $UBV$, and therefore the distance to an open cluster was estimated using
the ZAMS relation in the ($M_V,~ B-V$) diagram. As the earlier CCDs were
sensitive to longer wavelength, many $VRI$ or $BVRI$ CCD observations were
performed in the 1990s. Although the flux measured at longer wavelength is 
less sensitive to the stellar parameters, the $RI$ pass bands have their own 
merits. They are less affected by the interstellar reddening and less sensitive
to the difference in metallicity. The ZAMS relations in the optical pass bands,
especially the $U$ and $B$ filters, can be affected by the difference in metallicity \citep{sung99a}
as well as the chromospheric activity \citep{sung02}. 

\citet{sung04a} introduced a reddening-independent quantity Q$_{VI} \equiv V - 2.45 
(V-I)$ to estimate the distance to the starburst cluster NGC 3603. We introduce
four reddening-independent quantities Q$'$, Q$_{VJ}$, Q$_{VH}$, and Q$_{VK_s}$.

\begin{equation}
Q' \equiv (U-B) - 0.72 (B-V) - 0.25 E(B-V)^2
\end{equation}
\begin{equation}
Q_{VJ} \equiv V - 1.33 (V-J)
\end{equation}
\begin{equation}
Q_{VH} \equiv V - 1.17 (V-H)
\end{equation}
\begin{equation}
Q_{VK_s} \equiv V - 1.10 (V-K_s)
\end{equation}

The parameter Q$'$ is a modification of Johnson's Q to take into account
the effect of $E(B-V)$ on $E(U-B)/E(B-V)$, which is a non-negligible effect
for highly reddened stars. The other three parameters are derived from
the relation between $R_V$ and color excess ratios (see Section 5.1 or
\citet{guetter89}). We determine the distance to an open cluster in the
CMDs composed of Q$'$ and Q$_{VI}$, Q$_{VJ}$, Q$_{VH}$, or Q$_{VK_s}$ as shown
in Figure \ref{fig_dm}.

The advantages of these quantities are (i) they are independent of interstellar
reddening, (ii) homogeneous $JHK_s$ data are available from 2MASS, and 
(iii) they are less affected by differences in metallicity.
In most cases we determine the distance to the cluster using the data for O
and B-type stars, and so all the quantities are nearly free from metallicity
differences. In addition, we can check for differences or errors in the
photometric zero points from four CMDs, as shown in Figure \ref{fig_dm}.
The photometric zero point of \citet{park01} is well consistent with those
of 2MASS.

\subsection{Age and Initial Mass Function}

To determine the age and the IMF of an open cluster, we have to construct
the HRD of the cluster using the calibrations adopted in Section 4. 
In addition, we
need to adopt stellar evolution models to estimate the age and mass of 
a star. For a long time we have used the evolution models by the Geneva group
\citep{schaller92}. Recently, \citet{brott11} and \citet{ekstrom12} published 
new stellar evolution models with stellar rotation. The results (age and 
the IMF) are very similar to each other (see for details \citet{hur12,sung13}). 

As our survey is to observe a wide range of cluster ages, we will use the 
stellar evolution models by the Geneva group \citep{ekstrom12} because their 
models cover a wide range of initial masses.

The importance of PMS stars in young open clusters is that their age and mass 
can be determined using PMS evolutionary models. The age distribution of a cluster 
represents the star formation history in the cluster. In addition, because
many of our target clusters are relatively sparse, it is not easy to determine 
the age of a young open cluster with one or two O- or early B-type stars.
On the other hand, as the low-mass PMS stars are relatively rich, it is easy
to determine the age and the age spread of the cluster using low-mass PMS stars.
We determine the age and mass of PMS stars in young open
clusters using the PMS evolutionary models of \citet{siess00} since their
models cover a relatively large mass range.

\section{SUMMARY}

Open clusters are the most important objects to test observationally
stellar evolution theory which is a basic step toward an understanding of
the Universe. We started the Sejong Open cluster Survey (SOS) dedicated
to provide homogeneous photometry of a large number of open clusters in 
the Johnson-Cousins' $UBVI$ system which is tightly matched to the SAAO 
standard system. The goals of this survey project are to study the various 
aspects of star formation and stellar evolution, and the structure, formation 
history and evolution of the Galactic disk.

In this paper, we also described the target selection criteria
and the spatial distribution of targets based on the analysis of the open 
cluster database. In addition, we presented our strategy to achieve
accurate and precise photometry. 

To fulfill the goals of this project, we adopt and propose various calibrations
and tools required for the analysis of photometric data. We described
and compared various calibrations, and presented the adopted relations.
These include the intrinsic color relations between various colors, 
the zero-age main sequence relations, the spectral type versus absolute
magnitude relations, the spectral type versus effective temperature relations,
the spectral type versus color relations, the color - temperature relations,
and the temperature versus bolometric correction relations. In addition,
we presented methods to determine the reddening law $R_V$ and the distance
to the cluster.

\acknowledgments{ H.S. acknowledges the support of the National Research 
Foundation of Korea (NRF) funded by the Korea Government (MEST) 
(Grant No.20120005318). }



\end{document}